\documentclass[11pt]{article}
\usepackage{jcappub}
\usepackage{color}
\usepackage{amsfonts, amssymb,amsmath}
\usepackage{hyperref}
\usepackage{comment}
\usepackage{xspace} 

\usepackage{soul}

\definecolor{ForestGreen}{rgb}{0.13, 0.55, 0.13}
\definecolor{airforceblue}{rgb}{0.36, 0.54, 0.66}
\definecolor{orange}{rgb}{1.0, 0.5, 0.0}
\definecolor{amethyst}{rgb}{0.6, 0.4, 0.8}
\definecolor{awesome}{rgb}{1.0, 0.13, 0.32}
\definecolor{chromeyellow}{rgb}{1.0, 0.65, 0.0}
\definecolor{pigicolor}{rgb}{0.1, 0.6, 0.6}

\newcommand{\bx}{\mathbf{x}}
\newcommand{\bk}{\mathbf{k}}

\newcommand{\Mpc}{\, h^{-1} \, {\rm Mpc}}

\newcommand{\pinocchio}{\textsc{Pinocchio}\xspace}

\title{Fitting Covariance Matrix Models to Simulations}

\author[a,b,c,d,1]{Alessandra Fumagalli,\note{Corresponding author.}}
\author[c,e,b,d]{Matteo Biagetti,}
\author[a,b,c,d]{Alex Saro,}
\author[b,c,d]{Emiliano Sefusatti,}
\author[f]{An\v{z}e Slosar,}
\author[a,b,c,d]{Pierluigi Monaco,}
\author[g,h]{and Alfonso Veropalumbo}

\affiliation[a]{Astronomy Unit, Dipartimento di Fisica, Universit\`a di Trieste, via Tiepolo 11, 34143 Trieste, Italy}
\affiliation[b]{Istituto Nazionale di Astrofisica, Osservatorio Astronomico di Trieste, via Tiepolo 11, 34143 Trieste, Italy}
\affiliation[c]{Institute for Fundamental Physics of the Universe, via Beirut 2, 34151 Trieste, Italy}
\affiliation[d]{Istituto Nazionale di Fisica Nucleare, Sezione di Trieste,  via  Valerio  2,  34127 Trieste,  Italy}
\affiliation[e]{SISSA - International School for Advanced Studies, via Bonomea 265, 34136 Trieste,  Italy}
\affiliation[f]{Physics Department, Brookhaven National Laboratory, Upton NY 11973, USA }
\affiliation[g]{ Dipartimento di Fisica, Universit\`a degli Studi Roma Tre, via della Vasca Navale 84, 00146 Roma, Italy}
\affiliation[h]{Istituto Nazionale di Fisica Nucleare, Sezione di Roma Tre, via della Vasca Navale 84, 00146 Roma, Italy}

\emailAdd{alessandra.fumagalli@inaf.it}

\abstract{Data analysis in cosmology requires reliable covariance matrices. Covariance matrices derived from numerical simulations often require a very large number of realizations to be accurate. When a theoretical model for the covariance matrix exists, the parameters of the model can often be fit with many fewer simulations. We write a likelihood-based method for performing such a fit.  We demonstrate how a model covariance matrix can be tested by examining the appropriate $\chi^2$ distributions from simulations. We show that if model covariance has amplitude freedom, the expectation value of second moment of $\chi^2$ distribution with a wrong covariance matrix will always be larger than one using the true covariance matrix. By combining these steps together, we provide a way of producing reliable covariances without ever requiring running a large number of simulations. We demonstrate our method on two examples. First, we measure the two-point correlation function of halos from a large set of $10000$ mock halo catalogs. We build a model covariance with $2$ free parameters, which we fit using our procedure. The resulting best-fit model covariance obtained from just $100$ simulation realizations proves to be as reliable as the numerical covariance matrix built from the full $10000$ set. We also test our method on a setup where the covariance matrix is large by measuring the halo bispectrum for thousands of triangles for the same set of mocks. We build a block diagonal model covariance with $2$ free parameters as an improvement over the diagonal Gaussian covariance. Our model covariance passes the $\chi^2$ test only partially in this case, signaling that the model is insufficient even using free parameters, but significantly improves over the Gaussian one.}

\keywords{cosmological parameters from LSS, galaxy clustering,dark matter simulations, Bayesian reasoning}

\begin{document}
\setcounter{tocdepth}{2}
\maketitle

\thispagestyle{empty}
\setcounter{page}{2}


\def\thefootnote{\arabic{footnote}}
\setcounter{footnote}{0}

\baselineskip= 15pt

\newpage

\section{Introduction}

Modern studies of the large-scale structure rely on mapping the Universe using a variety of tracers and using the statistical properties of their distribution to put constraints on cosmological models. 
In most analyses, data are compressed into summary statistics that capture certain properties of cosmological perturbations, with examples including traditional measurements like power spectra and higher-order correlations \cite{BlakeEtal2011, DelaTorreEtal2013,AlamEtal2017, GilMarinEtal2017, AsgariEtal2021, AbbottEtal2022}. Theoretical and modeling progress have finally made it possible to include statistics beyond the power spectrum, like the bispectrum, in cosmological analyses of galaxy surveys \cite{GilMarinEtal2017, PhilcoxIvanov2022, DAmicoEtal2020}. Large collaborations, such as Euclid, are already preparing to exploit these statistics.  More complex observables have also been explored, such as various topological measures \cite{Biagetti:2020skr,Heydenreich:2020hrr,BiagettiEtal2022A,Heydenreich:2022dci}, synthetic tracers such as voids \cite{2010MNRAS.403.1392L} or galaxies marked by local density, morphology or stellar mass \cite{2006MNRAS.369...68S, 2016JCAP...11..057W, 2018MNRAS.478.3627A}, etc. This implies that data vectors will soon have hundreds of components. In all cases, however, it is necessary to obtain a robust estimate of the uncertainty affecting the measurement in order to derive solid constrains on the cosmological model. 

Relying on a numerical estimate based on simulations and mocks has been so far the most common approach, as it allows including more easily subtle observational effects (see, e.g. \cite{ManeraEtal2013, KitauraEtal2016, AvilaEtal2018}). 
For experiments with complex selection functions, one might be able to generate $\mathcal O(10)$ or $\mathcal O(100)$ realistic realizations, but it is unlikely that $10000$ would be feasible. For this purpose, several approximate, but efficient, methods to produce galaxy mock catalogs have been proposed over the past decade (see \cite{Monaco2016, FengEtal2016} for a review and a recent proposal or \cite{LippichEtal2019, BlotEtal2019, ColavincenzoEtal2019} for comparisons in terms of the predicted covariance).
On the other hand, upcoming galaxy surveys will pose a challenge to this method as mock galaxy catalogs will require high-resolution simulations over very large volumes and covering a very large parameter space, making their production extremely expensive when feasible at all. In this respect, over the last years, several works addressed the problem of a poor numerical estimate of the precision matrix \cite{HartlapEtal2009, TaylorJoachimiKitching2013, DodelsonSchneider2013, PercivalEtal2014, SellentinHeavens2016}, while various strategies to reduce the number of required realizations have been explored \cite{Scoccimarro2000B, HamiltonRimesScoccimarro2006, PopeSzapudi2008, Joachimi2017, PazSanchez2015, ChartierWandelt2022, deSanti:2022nlz}.

An alternative approach based on analytical models for the covariance matrix is also gaining attention \cite{Lacasa2020, FangEiflerKrause2020, SugiyamaEtal2020, WadekarScoccimarro2020, WadekarIvanovScoccimarro2020, BiagettiEtal2021A}.   
Analytical models can be affected by systematic theoretical errors and they still need to be validated against simulations. In particular, often such models depend on nuisance parameters to be determined by proper fits to numerical results. 

In this paper, we discuss a hybrid approach, which builds up on recent techniques, with the goal of obtaining covariance matrices that are simultaneously reliable and based on few simulations.
There are two insights our result is based on. The first one is that the covariance matrix estimation is, fundamentally, an estimate of two-point correlations of realizations of the data vector. This is a problem that has been solved many times in various fields, and for which an exact likelihood can be written and evaluated. It can be shown that the naively calculated numerical matrix contains all the information needed to evaluate the likelihood of a theoretical model of the same covariance matrix.
Previous work has exploited this insight to fit a model covariance to mock simulations for the Baryon Acoustic Oscillations (BAO) analysis of SDSS DR7 \cite{Xu:2012hg} and the two-point correlation function  \cite{OConnellEtal2016}, the bispectrum  \cite{Slepian:2015hca} and the power spectrum \cite{Pearson:2015gca} of BOSS data. In particular, \cite{OConnellEtal2016} and \cite{Pearson:2015gca} have used these fits to reduce the number of mocks needed to build a reliable covariance. 

The second insight is that, ultimately, any covariance matrix is used to compare some data with a model. The fundamental quantity of interest is therefore the $\chi^2$ distribution. A good covariance matrix  produces $\chi^2$--\,distributed values when presented with realizations of data vectors. A biased or inappropriate covariance matrix produces $\chi^2$ values that are not drawn from the correct distribution. This then tells us how to test a given covariance matrix. Applications of this idea have been investigated recently \cite{Hall:2018umb,DES:2020ypx}. 
Essentially, we propose to combine these two insights into a single algorithm to produce reliable and cheap covariance matrices.
Assuming we have $50$ realizations of a data vector of size 100, therefore corresponding to a covariance matrix of size $100\times 100$, we are unable to estimate the covariance with precision, but we can both i) test candidate covariance matrices (i.e. $50$ $\chi^2$ values are sufficient to show compatibility with the correct distribution) and ii) fit a few parameter model covariance matrices.


This paper is structured as follows. In Section \ref{sec:method} we present our method. We develop  two different tests to illustrate our method and show the results of the tests in Sections \ref{sec:test1} and \ref{sec:test2}. The last section presents conclusions and outlines future work.

\section{Covariance Matrix Estimation is Covariance Estimation}
\label{sec:method}
\newcommand{\vd}{\mathbf{d}}
\newcommand{\vm}{\mathbf{m}}
\newcommand{\vt}{\mathbf{t}}
\newcommand{\mC}{\mathrm{C}}
\newcommand{\mX}{\mathrm{X}}
\newcommand{\mCt}{\mathrm{C}_{t}}
\newcommand{\rT}{\mathrm{T}}
\newcommand{\Tr}{{\rm Tr}}
\newcommand{\M}{N_{\rm sims}}
\newcommand{\Var}{\mathrm{Var}}
\newcommand{\Mean}{\mathrm{Mean}}

 Let us assume we are interested in the covariance matrix for a quantity that can be represented by an $N$ dimensional vector. An example might be a measurement of the power spectrum of galaxies in $N$ bins, or the measurement of a bispectrum, or both. 
 
Ultimately, we want to run a Bayesian analysis, comparing a measurement vector $\vm$ (of size N) with a theory prediction $\vt(\theta_m)$, where $\theta_m$ are the parameters of the model for the \emph{expectation value} of $\vm$. Assuming a Gaussian likelihood, we have
\begin{equation}
    P(\theta_m | \vm) \propto P(\vm | \theta_m) = (2\pi)^{-N/2}\,\left| \mC \right|^{-1/2} \exp \left\{-\frac{1}{2} \left[\vm - \vt(\theta_m)\right]^\rT \mC^{-1} \left[\vm - \vt(\theta_m)\right] \right\} \label{eq:meanlike}\,.
\end{equation}
In principle, the matrix $\mC$ depends as well on the parameters $\theta_m$. If  such dependence is significant, our method can be straightforwardly extended to accommodate it as well. For now, we assume that this is not the case.

Suppose we have $\M$ simulations providing as many  realizations of the measurements $\vm_i$, where $i=1\ldots \M$. We assume that these simulations are drawn from the same underlying theory and that the output varies only because of a different realization of the cosmic structure and noise. In practical terms, they are identical runs except for the random seed. By construction, the ensemble mean value of those realizations is $\left<m_i\right>\equiv\vm(\theta_{m,{\rm sim}})$  where $\theta_{m,{\rm sim}}$ are the fiducial values of the parameters adopted  to create the  simulations. These are assumed to be the same for all realizations.  Defining $\vd_i = \vm_i - \vm(\theta_{m,{\rm sim}})$, we see that the $d_i$ are normally distributed around zero with covariance $\mC$.\footnote{Note that in this paper, as is often the  practice, we subtract the mean of simulation results rather than the theoretical true model. For normally distributed residuals and an accurate theoretical model this is equivalent given a sufficient number of simulations.  If the difference is significant, one has a bigger fish to fry first before worrying about the covariance matrix.}

We can  write the naive matrix estimator
\begin{equation}\label{eq:defnumcov}
    \mC_n = \frac{1}{\M} \sum_{i=1\ldots \M} \vd_i \vd_i^\rT.
\end{equation}
We refer to this covariance matrix  as the ``numerical'' covariance matrix. 

Now, let us assume that we have a theoretical model for $\mC$ depending on some model parameters $\theta$, i.e. $\mC = \mC(\theta)$. That is, for a given set of $\theta$s, we can predict all values of $\mC$. Note that parameters $\theta$ cannot contain $\theta_m$, since we have just assumed that these do not affect $\mC$. 

The model for $\mC$ can be either physical, based on theoretical expectation about $\mC$, but it can also be purely phenomenological, e.g. assuming $\mC$ has a Toeplitz form, or that the off-diagonal terms beyond the second diagonal vanish. In any case, we can now use  the Bayesian theorem to write a likelihood for the covariance parameters $\theta$ as:
\begin{equation}
    P(\theta | \vd ) \propto P(\vd | \theta) \Pi(\theta),
\end{equation}
where we can put any prior information in $\Pi(\theta)$ and which we assume to be unity and where 
\begin{equation}
\mathcal{L} = P(\vd | \theta) \  \propto \prod_{i=1\ldots \M} |\mC(\theta)|^{-1/2} \exp\left[-\frac{1}{2}\vd_i^\rT \mC^{-1}(\theta) \vd_i\right]\,.
\end{equation}

For the classically educated cosmologist, this equation looks quite familiar. It is the equation representing the likelihood of a Gaussian field. 
The log-likelihood function $L \equiv \log \mathcal{L}$ equals up to a constant to
\begin{equation}
    L(\vd |  \theta ) \, =  - \frac{\M}{2} \log |\mC(\theta)| -\frac{1}{2}\sum_i \vd_i^\rT \mC^{-1}(\theta) \vd_i = - \frac{\M}{2} \left[\log |C(\theta)| + \Tr (\mC^{-1}(\theta) \mC_n)\right] \,.
    \label{eq:master}
\end{equation}
This is the form of the equation that we use in what follows.  This equation appears in this form already in \cite{OConnellEtal2016,Slepian:2015hca}, but in the context of very concrete models for the covariance matrix.  In this paper, we stress that this form is general and that it can be applied to any model for covariance matrix and for observables beyond 2-point correlation function, as we will do in Section \ref{sec:test2}. The most interesting aspect is that the likelihood can be rewritten in a form that depends only on the numerical covariance matrix $\mC_n$. In other words, we can compress the results of $N_{\rm sims} $ simulations into a single $N\times N$ matrix. If $N_{\rm sims}  > N$ this form offers useful information compression. It is also true if $N_{\rm sims} <N$, i.e. $\mC_n$ can in principle be even non-invertible and still contain all the available information at the 2-point level from the simulation suite. 
Given a sufficient number of simulations that the uncertainty on $\theta$ is negligible compared to the measurement noise on $\vm$, it suffices to find the maximum likelihood point in the $\theta$ space and use the resulting matrix. 

In what follows, we demonstrate this technique in practice with two examples.

\section{Goodness of fit for covariance matrix model}
\label{sec:chi2test}

After deriving the best-fit parameters for a given model of the covariance matrix, we would like to determine if the result is indeed a useful covariance matrix. Note that this is not a model comparison exercise, but a problem of goodness-of-fit: is the resulting covariance matrix actually fit for purpose?

A common use of the covariance matrix is the evaluation of the $\chi^2$ for a cosmological model likelihood. Under the assumption of Gaussianity, a good covariance matrix is the one providing correctly distributed $\chi^2$ values. A simple test on the inverse can be done by verifying that the residuals $\vd_i=\vm_i-\vm(\theta_{m,{\rm sim}})$ are $\chi^2$--\,distributed with the right number of degrees of freedom,
\begin{equation}
\chi_{th,i}^2 =  [\vm_i-\vm(\theta_{m,{\rm sim}})]\,\mC^{-1}\, [\vm_i-\vm(\theta_{m,{\rm sim}})]^T.\,
\end{equation}
While one would ideally have a separate set of simulations, this test can also be performed on the \emph{same} simulations that were used to infer the parameters. This is equivalent to fitting a theory to data and then using the same data to check the resulting $\chi^2$ \emph{without subtracting the model degrees of freedom}. This is a valid procedure where the number of simulations is much larger than the number of free parameters in the theory. When this condition is violated, the only safe thing to do is to split simulations into a "training" and "testing" subsets.

We will now consider how do the moments of $\chi^2$ distribution respond to being tested with a wrong covariance matrix. First consider a simple model where we simply fit for the covariance matrix amplitude, i.e. 
\begin{equation}
    C(\theta_A) = \theta_A C_0.
\end{equation}
At this point, we make no claims of whether $C_0$ is a good or poor approximation of the true covariance matrix, it is simply a matrix.
We have $|C(\theta_A)| = \theta_A^N |C_0|$ and $C^{-1}(\theta_A) = \theta_A^{-1} C_0^{-1}$. Plugging these expressions into Equation \ref{eq:master} we find that the maximum likelihood point is given by
\begin{equation}
\theta_A = \frac{\Tr (\mC^{-1}_0 \mC_n)}{N}.
\label{eq:solA}
\end{equation}
For the simulation realization $i$, the $\chi^2$ is given by $\chi^2_i = \vd_i^TC^{-1}\vd$ (for some covariance matrix $C$) and so the mean over the set of simulations is given by
\begin{eqnarray}
    \Mean\ \chi^2 &=&  \frac{1}{\M} \sum_{i=1}^{\M} \chi^2_i =  \Tr (\mC^{-1}(\theta) \mC_n) = N, \label{ref:firstm}\\
    \Var\ \chi^2 &=& \frac{1}{\M} \sum_{i}^{\M} (\chi^2_i)^2 - \left( \Mean\ \chi^2 \right)^2 = \sum_{i}^{\M} (\vd_i^TC^{-1}\vd)^2 -N^2 ,
\end{eqnarray}
The last equality of equation \ref{ref:firstm} comes from using solution of Equation \ref{eq:solA} and doing some straightforward manipulation.  Even if $C_0$ is a completely wrong, the terms containing trace of $\mC_0^{-1} \mC_n$ cancel exactly and so one always has $\Mean\ \chi^2=N$. In other words, if our model for covariance matrix has a freedom to adjust the amplitude, then the maximum likelihood is such that the first moment of $\chi^2$ matches the theoretical expectation.
Note that in general, our model for $C$ will have many more parameters, but as long as there is a subspace of the model which corresponds to a simple amplitude rescaling, this statement will be true.

\newcommand{\mCT}{\mC_{\rm true}}

There is no such simplification for variance. In this case we can calculate the expectation value and find after some simplifications that:
\begin{eqnarray}
    \left< \Var\ \chi^2 \right> &=& 2 \Tr (\mC^{-1}(\theta) \mCT \mC^{-1}(\theta) \mCT),
\end{eqnarray}
where $\mCT$ is the true covariance matrix, i.e. the one from which vectors $\vd$ are drawn. If $\mC(\theta)=\mCT$, i.e. if our model covariance matrix is indeed true, we find the standard moment of $\chi^2$ distribution, i.e. $\left <\Var\ \chi^2 \right> = 2 \Tr (\mathrm{I}) = 2N $.

Next let us assume $\mC(\theta)$ is different from $\mCT$, so that its inverse is given by  $\mC^{-1}(\theta)=\mCT^{-1}+\mX$. Additionally, let us assume that the first moment is correctly recovered as will always happen when the model has the freedom to rescale the matrix. In this case we find that 
\begin{equation}
 \Tr (\mCT \mX) = \Tr (\mCT C^{-1}(\theta)) - \Tr (\mCT \mCT^{-1}) = \left<\chi^2\right> - N =  0.
\end{equation}
Therefore
\begin{eqnarray}
        \left< \Var\ \chi^2 \right> &=& 2N + 2 \rm \Tr(\mCT \mX \mCT \mX) > 2N
\end{eqnarray}
To show the second line inequality, we note that $\mCT$ is a positive definite matrix and we can always rotate it into the frame $\mCT'$ where it is diagonal with its eigen values $\lambda_i$ on the diagonal. In this frame $\mX$ becomes $\mX'$ and the second term becomes $2 \rm \Tr (\mCT' \mX' \mCT' \mX') = \sum_{ij} \lambda_i \lambda_j \mX'^{2}_{ij} > 0.$

Finally, let's consider the error on the $\chi^2$ on a given data vector, which appears as a result of using the wrong covariance matrix. For a given $\vd$
\begin{equation}
    \Delta \chi^2 (\vd)  = \vd^T\mC(\theta)^{-1} \vd - \vd^T\mC^{-1}\vd = \vd^T \mX \vd
\end{equation}
Using the same assumption as above it is easy to show that the first and second moment of this quantity are
\begin{eqnarray}
        \left< \Delta \chi^2 \right>  &=& 0 \\
        \left< \Var\  \Delta \chi^2 \right>   &=& 2 \rm \Tr(\mC \mX \mC \mX) \label{eq:chi2scat}
\end{eqnarray}
In other words, for data vectors drawn from the true covariance matrix, using the wrong covariance matrix produces additional scatter around true $\chi^2$ values. This additional scatter has zero mean and variance given by Eq. \ref{eq:chi2scat}. Since variance add, this results in a distribution of $\chi^2$ values that is broader by the same amount. 

To recap, we have shown three simple but powerful results. If the model covariance matrix has freedom to vary in amplitude, the maximum likelihood will adjust its value so that the expectation value of $\chi^2$ is the correct value. In that limit, the values of individual $\chi^2$ computed with a wrong covariance matrix scatter around their true values: some of them are lower and some of them are higher than what they should be. Because variances add, this results in a
second moment of $\chi^2$ distribution that is larger that $2N$, with equality holding when $\mC(\theta)$ is the correct covariance matrix.

This gives us a direct handle on accuracy of covariance matrix. If we are unwilling to tolerate more than $\Delta \chi^2=1$ in $\chi^2$ error, then we should really find a model in which both the first and the second moment are reproduced to this accuracy. In practice, we might find that a significantly more relaxed $\chi^2$ can still produce essentially unchanged constraints.

\section{Test 1: Two-Point Correlation function}
\label{sec:test1}

For the first test of our method, we consider the two-point correlation function as observable, measured from a very large number $N_{\rm sims} =10000$ of mock halo catalogs. Let us assume as original problem a test of the two-point correlation function model based on the correct recovery  of the cosmological parameters of the simulations. For the sake of simplicity, we only consider $\Omega_m$ and $\sigma_8$. Such test requires the knowledge of the covariance matrix. Given the large number of mocks, we can build a reliable numerical covariance matrix, $\mC_n$. Using the method presented in Section \ref{sec:method}, we show how we can obtain an equally reliable covariance matrix using only a fraction of the $N_{\rm sims} $ simulations available.

\subsection{The two-point correlation function and its covariance}
The basic ingredients of this test involve a model for the two-point correlation function of halos corresponding to $\mathbf{t}(\theta_m)$  in the previous section and the model for covariance matrix corresponding to $\mC(\theta)$.

For the purpose of testing our covariance fitting technique, we take a simple linear model for the halo power spectrum including shot noise,
\begin{equation}\label{eq:pk}
P_h(k) = b^2 P_m(k) + \frac{1+\alpha}{\bar n},
\end{equation}
where $b$ is the linear bias, $P_m$ is the linear matter power spectrum, $\bar n$ is the halo number density and $\alpha$ parametrizes deviations from Poissonian shot noise, $1/\bar n$. The corresponding two-point correlation function is given by,
\begin{equation}\label{eq:zk}
\xi_h(r) =  b^2 \xi_m(r),
\end{equation}
where $\xi_m$ is the Fourier transform of $P_m$. When comparing to simulations, we fit the linear bias to the measurements of the two-point correlation function. 
In order to compute the covariance for Eq.\,\eqref{eq:zk}, we Fourier transform the leading\footnote{The full covariance would include a trispectrum term \cite{ScoccimarroZaldarriagaHui1999}, which we assume to be negligible \cite{GriebEtal2016}.} term of the power spectrum covariance
\begin{equation}\label{eq:model}
    \mC(b,\alpha) \simeq \frac{2}{V} \int \frac{{\rm d} k \, k^2}{2 \pi^2}  \bigg ( b^2\,P_m(k) + \frac{(1+\alpha)}{\bar{n}} \bigg )^2 W_i(k) W_j(k) \,, 
\end{equation}
where $V$ is the box volume and $W_i(k)$ is the Fourier transform of the i-th radial shell
\begin{equation}
    W_i(k) = \int \frac{{\rm d}^3r}{V_i} j_0(kr) = \frac{r_{i,+}^3 W_{\rm th}(k r_{i,+}) - r_{i,-}^3 W_{\rm th}(k r_{i,-})}{r_{i,+}^3 - r_{i,-}^3} \,,
\end{equation}
where $V_i = \dfrac{4 \pi}{3} \left(r_{i,+}^3 \, - \, r_{i,-}^3\right)$ is the shell volume, $j_0(kr)$ is the zero-th order spherical Bessel function, and $W_{\rm th}(kr)$ is the Fourier-transform of the top-hat window function. Our model covariance therefore depends on two parameters $\theta = \{ b,\alpha\}$. A simple choice for these parameters would be to use the bias we fit to the two-point correlation function measurements $ b \equiv \hat b$ and $\alpha=0$.

\subsection{Description of the Data}

We take advantage of a very large set of $N_{\rm sims} =10000$ mock halo catalogs obtained with the \pinocchio code \cite{MonacoTheunsTaffoni2002, MonacoEtal2013, MunariEtal2017}. The code provides an approximate dark matter halo distribution based on particle displacements computed from 3rd order Lagrangian Perturbation Theory and a halo identification from ellipsoidal collapse. In comparison with full N-body numerical simulations the latest version of the code is able to recover the halo mass function and linear halo bias of Friends-of-Friend halos with an accuracy respectively within 5\% and 10\% of the full N-body. 

The mock catalogs are built from $1000^3$ dark matter particles in a cubic box of side  $L=1500\Mpc$, reproducing the set-up and adopting the cosmology of the Minerva \textit{N}-body simulations \cite{GriebEtal2016}. The mass threshold is defined requiring that the large-scale amplitude of the \pinocchio catalogs power spectrum (including shot-noise) matches the amplitude of the same power spectrum measured in the Minerva simulations catalogs, the latter defined by a minimal mass of $M\simeq 1.12\times 10^{13}\; h^{-1} {\rm M}_{\odot}$. We refer the reader to  \cite{OddoEtal2020} for a more detailed description of the mock halo catalog construction, that allows to reproduce the power spectrum and bispectrum variance within 10\% \cite{OddoEtal2020, OddoEtal2021}.
The 2-point correlation function is measured using the standard Landy-Szalay estimator \cite{LandySzalay1993} as implemented in the \textsc{CosmoBolognaLib} code \cite{MarulliVeropalumboMoresco2016}, adapted, however, to account for the box periodic boundary conditions in the pair counting.

The numerical covariance  is estimated directly from the \pinocchio mocks as
\begin{equation}\label{eq:numcov}
    \mC_n(r_i,r_j) = \left\langle (\hat{\xi}_h(r_i) - \bar \xi_h(r_i))(\hat{\xi}_h(r_j) - \bar \xi_h(r_j))\right\rangle,
\end{equation}
where $\hat{\xi}_h(r_i)$ is the measurement from a single mock and $\bar \xi_h$ is the mean value, and the brackets indicate the average over the $N_{\rm sims} =10000$ mocks. The numerical covariance obtained from Eq.\,\eqref{eq:numcov}
has been compared with full simulations results, showing good agreement (but with a different definition of the mass threshold) in \cite{LippichEtal2019}. 

We should notice that the covariance model above does not account for the specific geometry of our distribution, that is the effect of the exact number of pairs in the case of a box with periodic boundary conditions \cite{PhilcoxEisenstein2019, LiEtal2019}. We therefore expect that the recovered value of the covariance model parameters accounts in part also for this neglected effect.

\begin{figure}[t]
    \centering
    \includegraphics[width=1.02\textwidth]{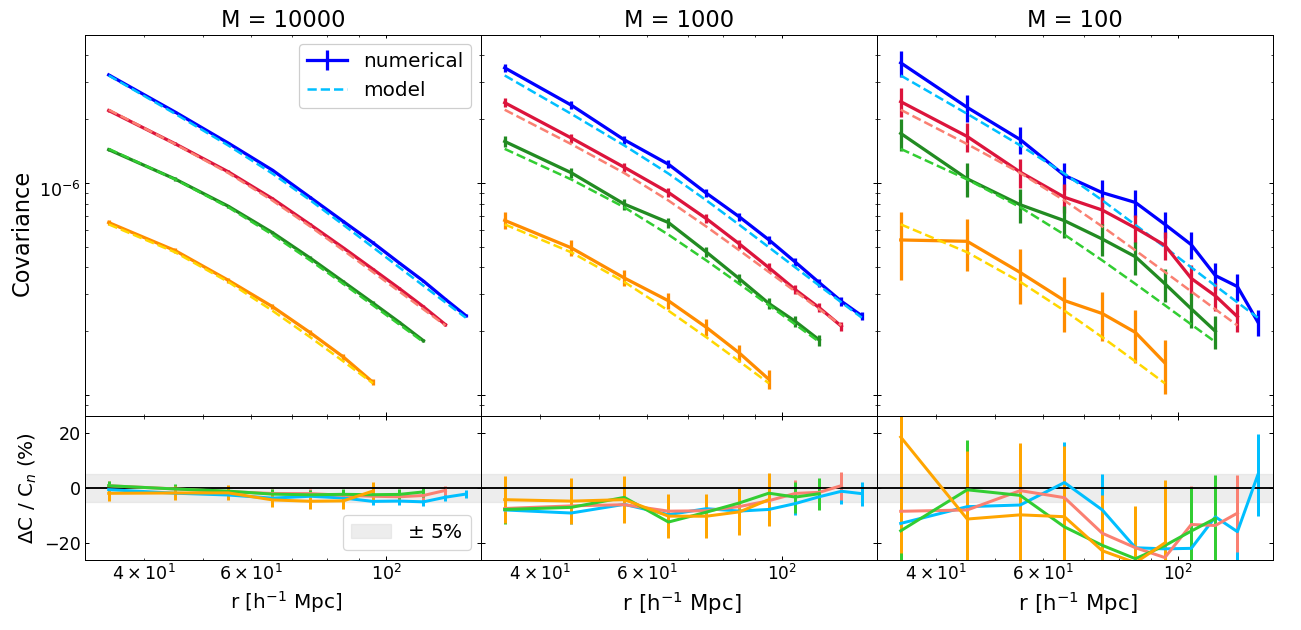}   
    \caption{Comparison between the model $C_{\xi}$ (dashed lines) and numerical (solid lines) covariance, computed with different number $N_{\rm sims} $ of simulations. Different colors represent different terms of the matrices: diagonal in blue, first off-diagonal in red, second off-diagonal in green and fourth off-diagonal in orange. In the bottom panels, percent residuals between numerical and model covariances. Errorbars are heavily correlated, which makes the scatter between neighboring points less than that implied by errorbars.}
    \label{fig_comp_numerical}
\end{figure}
\subsection{Results}
\label{sec:results}

In this section we outline the results of this test. The goal is to show that our model covariance is as reliable as the numerical one obtained from a very large set of simulations. The final test is therefore to demonstrate that cosmological parameters, in our case $\Omega_m$ and $\sigma_8$, are recovered with the correct value and uncertainty using the model covariance. The procedure can be summarized as follows:
\begin{enumerate}
    \item Build a set of numerical covariance matrices using Eq.\,\eqref{eq:numcov} for varying number of simulations $N_{\rm sims} =10000$, $1000$, $100$, $50$, and $30$.
    \item Build a model covariance with free parameters, following Eq.\,\eqref{eq:model}. The free parameters are the linear bias $b$ and the deviation from Poisson shot noise, $\alpha$.
    \item Maximize the likelihood of Eq.\,\eqref{eq:master} varying the free parameters, using the sets of $\mC_n$ from point 1) to get a best-fit model covariance $\mC(b_{\rm fit}, \alpha_{\rm fit})$.
    \item Verify the reliability of the model covariance matrix, and of all the numerical covariance matrices, for varying number of simulations.  
\end{enumerate}
We proceed then with details on the maximization procedure and the check of reliability of the model covariance. We also show results obtained by fixing the covariance parameters to the fit to the two-point correlation function, $\mC_\xi$.

\subsubsection{Maximizing the likelihood}

We fit our model covariance $\mC(b,\alpha)$ using the technique described in Section \ref{sec:method}. We perform the inference for $b$ and $\alpha$ using the likelihood of Eq.~(\ref{eq:master}) with the Nested Sampling Monte Carlo library \textsc{PyMultiNest} \cite{BuchnerEtal2014}. We assume flat uninformative priors for the two parameters, $b = [0, 5]$ and $\alpha = [-1, 1]$. We repeat the fit using several subsets of simulations $N_{\rm sims} =10000$, $1000$, $100$, $50$, $30$ to build the numerical covariance $\mC_n$. 
In Fig.\,\ref{fig_comp_numerical}, we show a comparison between the numerical and the model covariance matrices for three different subsets of simulations, $N_{\rm sims} =10000$, $N_{\rm sims} =1000$ and $N_{\rm sims} =100$. The errorbars are given by the standard deviation computed on a set of $n = N_{\rm sims, tot}/N_{\rm sims}$ independent subsets of simulations, where $\mathbf{N_{\rm sims,tot} = 10000}$ is the total number of available simulations. In this case, we fix the model covariance to be $C_\xi$, i.e. fixing $b$ to the value fit to the 2-point function measurements and $\alpha=0$. As we decrease the number of simulations, the model deviates from the numerical covariance. 

\begin{figure}
    \centering
    \includegraphics[width=0.53\textwidth]{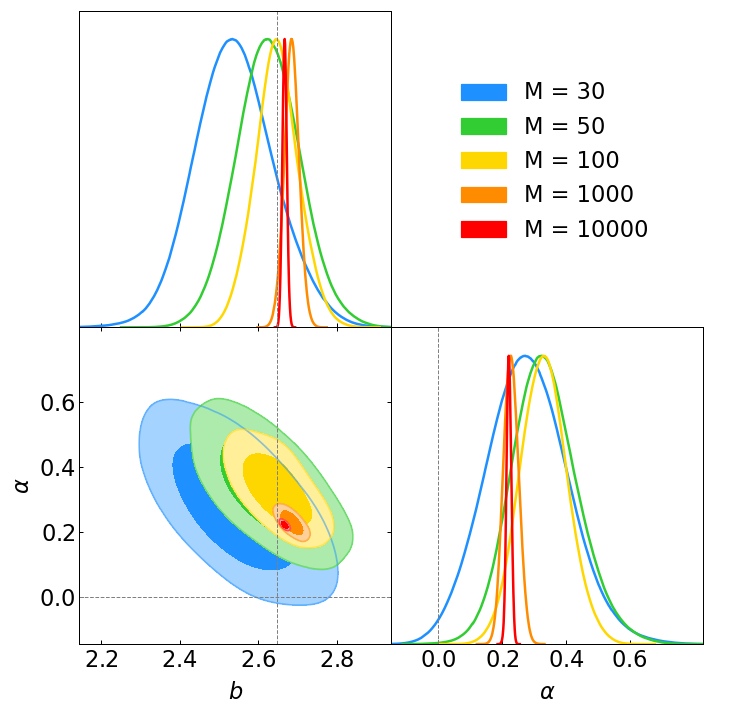}  
    \includegraphics[width=0.455\textwidth]{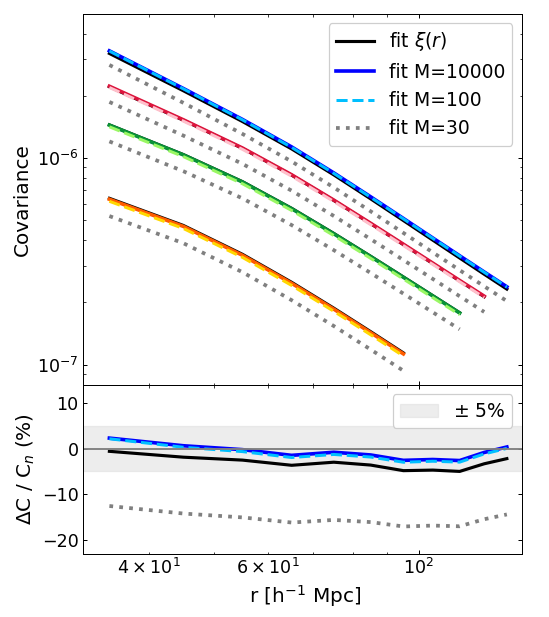}   
    \caption{\emph{Left Panel.} Contour  plots  at  $68$  and  $95$  per  cent  of  confidence  level  for the covariance parameters, fitted over different numbers of simulations. The gray dashed lines represent the reference values, given by $\alpha = 0$ (Poissonian shot-noise) and bias fitted from the 2-point correlation function. \emph{Right Panel.} Model covariances for the cases $N_{\rm sims} =10000$, $100$, $30$ and $\xi(r)$ fit (respectively solid dark, dashed light, dotted gray and black lines). In the bottom panel, percent residuals  with respect to the full numerical matrix. Color code as in Fig.\,\ref{fig_comp_numerical}.}
    \label{fig_fit}
\end{figure}

It is useful to show the posteriors resulting from the maximization of the likelihood, in the left panel of Fig.\,\ref{fig_fit}. As expected, as we use an increasing number of simulations, the contours shrink, but they are consistent with each other. Even with only $30$ simulations we can broadly constrain the parameters well within the prior. Interestingly, the data prefer a non-zero value of the shot noise parameter, and a value of the bias that disagrees with the one we have fit to the two-point correlation function (dashed line). This effect is distinguishable even with a very small number of simulations and it becomes very significant at $N_{\rm sims} =10000$ with an over 30 sigma tension between the standard Poisson shot noise and our measurement of it. As mentioned, this is not necessarily reflecting a proper departure from the Poisson limit of the halo distribution shot-noise, but it can include additional systematics, e.g. geometry effects, and the incompleteness of the model.\footnote{As a further check, we have fit $b$ and $\alpha$ directly on power spectrum measurements performed on the same set of simulations. The fitted values are also in disagreement with our best-fit values found using our model covariance. This confirms that the discrepancy is explained by the incompleteness of the model covariance, rather than a true deviation from Poisson shot-noise.} Another interesting comment is that, although the terms containing $\alpha$ are always subdominant, they contribute to the fitting process in a non-negligible way and with approximately the same weight of the main term. Thus, even if the difference between $C_\xi$ and $C_n$ is only about 5\,\%, it is not unreasonable for the fit to prefer a value of $\alpha$ with about a 20\,\% boost over the fiducial value ($\alpha \sim$ 0.2 instead of $\alpha=0$).

\begin{figure}
    \centering
    \includegraphics[width=0.99\textwidth]{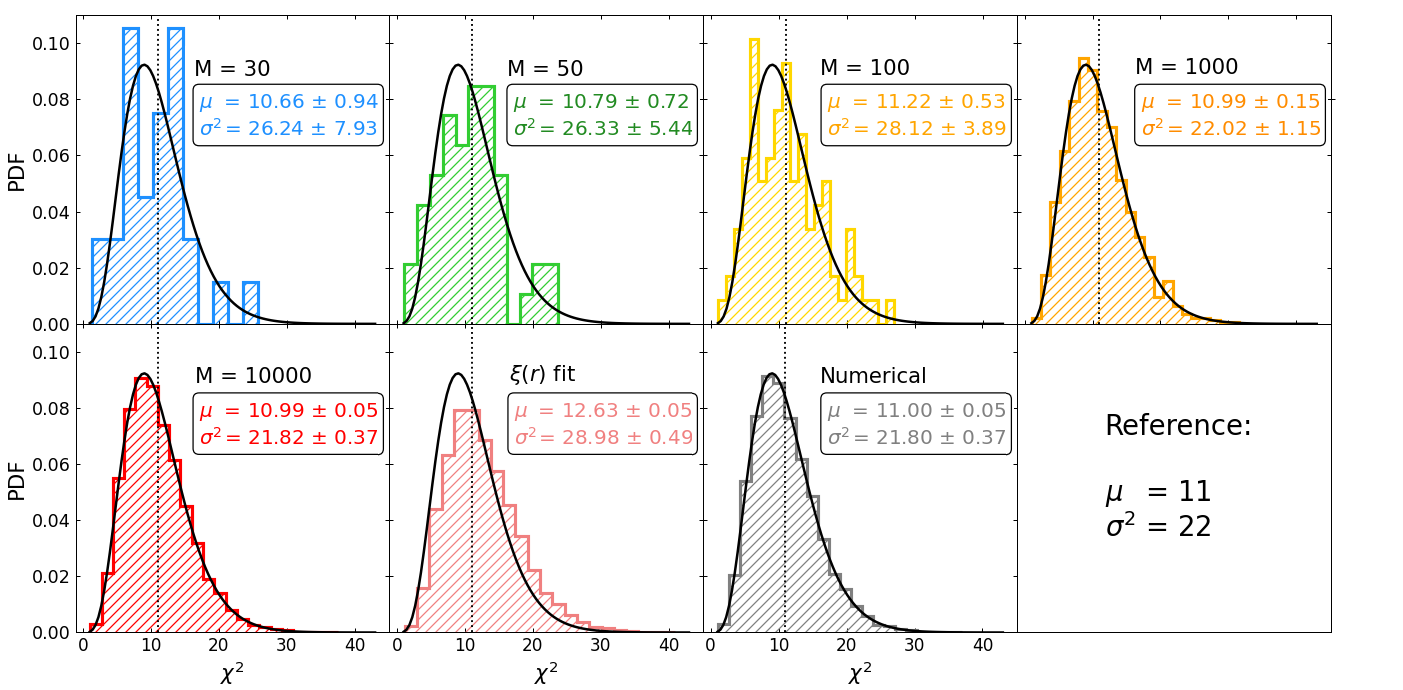}   
    \caption{$\chi^2$ distribution corresponding to the fitted parameters for different numbers of simulations (colored histograms), compared to the predicted distribution (black). In the bottom panels, comparison with the matrix with parameters fitted from $\xi(r)$  and the numerical matrix. The errors on the mean and the variance are computed with bootstrap method.}
    \label{fig_chi2}
\end{figure}
\subsubsection{A $\chi^2$ test for the inverse of the covariance}

In Fig.\,\ref{fig_chi2} we plot the histogram of $\chi^2$ values for the model covariances built from the subsets of simulations as described in Section \ref{sec:chi2test}. As a reference, we also calculate the histogram for the numerical covariance matrix using $N_{\rm sims} =10000$ simulations and the model covariance where we fix the bias from the correlation function fit and $\alpha=0$, $\mC_{\xi}$. All these candidate models for the covariance are compared to the theoretical $\chi^2$ distribution for this setup, which has mean $\mu=11$ and variance $\sigma^2=22$. We calculate the mean and variance of each histogram; their errors are estimated by bootstrap, i.e. by resampling the data and evaluating the confidence region from the bootstrap distribution of the resampled statistics. From this test it is clear that our model covariance agrees reasonably well with the theoretical curve already with only $N_{\rm sims} =30$ simulations, even though with large error bars. On the other hand, the model covariance with fixed parameters is significantly away from the predicted distribution. 

\subsubsection{Effects on cosmological parameters}
\begin{figure}
    \centering
    \includegraphics[width=0.49\textwidth]{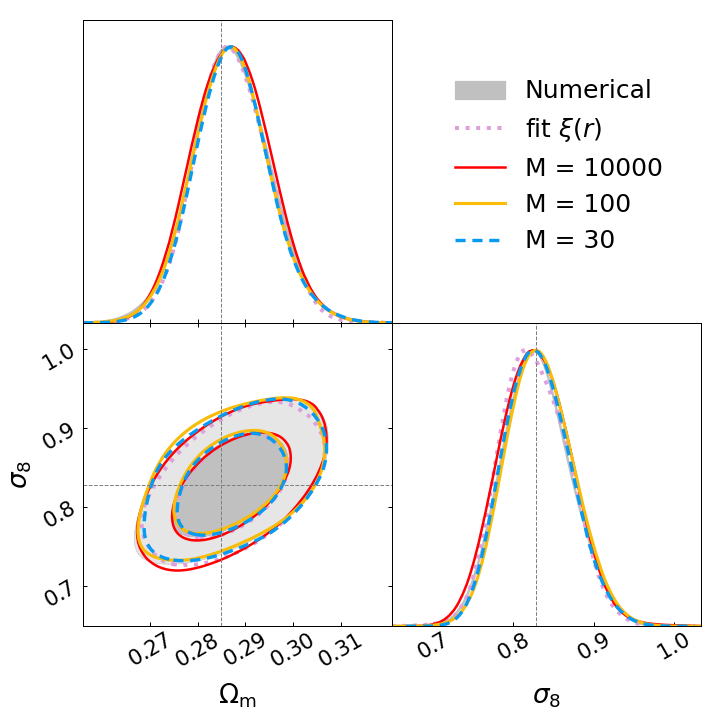} 
    \includegraphics[width=0.495\textwidth]{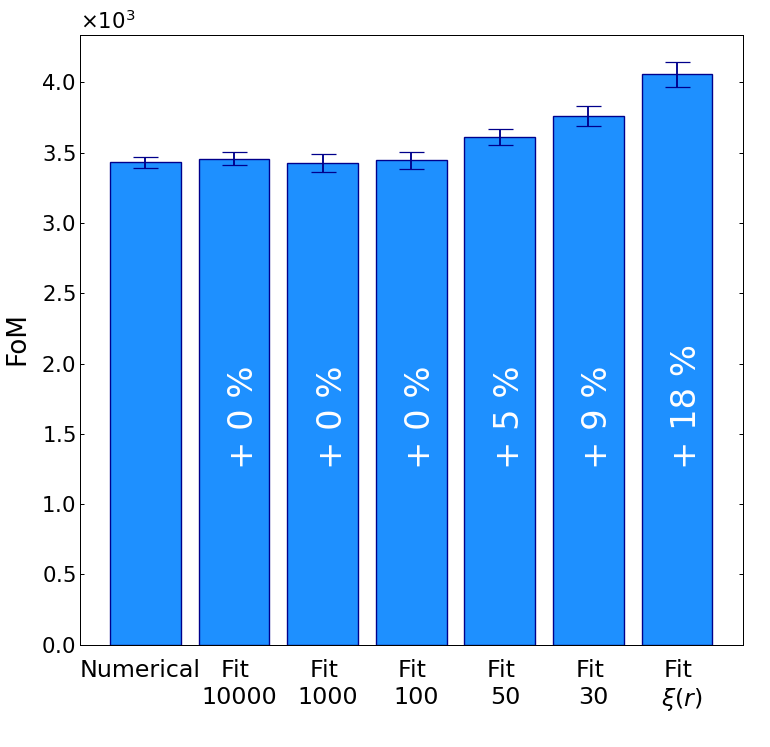}
    \caption{\emph{Left Panel:} Contour  plots  at  68  and  95  per  cent  of  confidence  level  for the cosmological parameters $\Omega_{\rm m}$ and $\sigma_8$, computed with different covariance matrices: full numerical (gray), no-fit model (pink), $N_{\rm sims} =30$ fit (blue), $N_{\rm sims} =100$ fit (yellow), $N_{\rm sims} =10000$ fit (red). \emph{Right Panel:} Figure of merit in the $\Omega_{\rm m} - \sigma_8$ plane for different covariance cases: the numerical covariance matrix drawn from the full set of $N_{\rm sims} =10000$ simulation, our model covariance for varying $N_{\rm sims} $ where we use the best-fit values for $b$ and $\alpha$ as found from the likelihood minimization, and the model covariance where we use $b$ fit from the two-point correlation function and $\alpha=0$. The error bars are computed from an average of $n = 10$ realizations, as $\epsilon = \sigma / \sqrt{n}$.}
    \label{fig_6}
\end{figure}
The final test is to show that from the clustering analysis of the two-point correlation function we are able to correctly recover cosmological parameters from the simulations by using our model covariance.
We consider a Gaussian likelihood for the observable, the two-point correlation function of halos. The theoretical model is the one given in Eq.\,\eqref{eq:zk}, where we vary the linear bias and two cosmological parameters, $\Omega_{\rm m}$ and $\sigma_8$, which enter through the linear matter power spectrum. We choose the following flat priors for these parameters: $\{b,\Omega_{\rm m},\sigma_8\} = \{[0,5], [0.20,0.35],[0.7,0.9]\}$. As for the covariance, we use our model covariance set to the best-fit values of Fig.\,\ref{fig_fit}. Note that we allow the bias of the covariance model to be different to the bias in the correlation function model. This is because we expect the former to also play a role in absorbing the missing terms in the covariance model, and thus lose its physical meaning of bias. For comparison, we also consider the case in which covariance and correlation function biases are the same, i.e. $\mC_\xi$. All the cases are compared to the results with the full numerical covariance $\mC_n$. The results are shown in the left panel of Fig.\,\ref{fig_6}. As expected, our model covariance with only $N_{\rm sims} =30$ simulations does not bias the contours of the parameters with respect to the numerical one built from $N_{\rm sims} =10000$ simulations. By eye, the fixed model covariance seems to be as good as the others, deviating by a small amount from the numerical one.
To better evaluate the differences between the results, we estimate the accuracy of the parameter posteriors by computing the figure of merit \cite{AlbrechtEtal2006a} in the $\Omega_{\rm m}$ - $\sigma_8$ plane
\begin{equation}
\mathrm{FoM}(\Omega_{\rm m }, \sigma_8) = \frac{1}{\sqrt{\det \left [ \mathrm{Cov}(\Omega_{\rm m }, \sigma_8) \right ] }} \,,
\end{equation}
where $C(\Omega_{\rm m }, \sigma_8)$ is the parameter covariance computed from the sampled points. To take into account the statistical uncertainty of the likelihood maximization process, we compute the figure of merit as the average over $n=10$ realizations, with errors given by the standard error $\varepsilon = \sigma/\sqrt{n}$. Clearly, the figure of merit here is to be intended as an overall estimate of the parameters errors and it is relevant only in its departure from the fiducial value. A good covariance matrix would give the same FoM as numerical fit on 10,000 realizations: a higher FoM indicates errors that are too small, while a low FoM indicates errors that are too large.

The result is shown in the right panel of Fig.\,\ref{fig_6}. It confirms that our model covariance works well already with $N_{\rm sims} =30$, and provides a perfect match for $N_{\rm sims} =100$ simulations.  We also find a curious effect: using too few simulations provides a covariance matrix that not only varies from one set of 30 simulations to another, but one that is also systematically biased towards high FoM (underestimation of the errorbars) . With too few simulations, the numerical matrix turns out to be inaccurate, providing a biased fit of the model. As shown in left panel of Fig.~\ref{fig_fit}, the covariance fitted from 30 simulations underestimates the true covariance; different results can be obtained from different subsets of simulations, as the data on which performing the fit can be biased in the opposite direction. The inaccuracy of such fit is confirmed by the FoM of Fig.~\ref{fig_6}.

Finally, we see that using a covariance where we fix the bias to the fitted value to the two-point correlation function measurements and $\alpha=0$, $C_\xi$, provides a $\sim 20\%$ underestimation of the error with respect to the numerical one.

\section{Test 2: Bispectrum}\label{sec:test2}

As a second test, we study the bispectrum, i.e. the three-point correlation function in Fourier space, on the same halo catalogs used for the first test. The motivation for choosing this observable as a test is twofold: first, it is an example where the covariance matrix can have very large dimensions. As compared to the power spectrum, where the covariance matrix is built from a data vector with typically $N \sim 20-50$ components, for the bispectrum the data vector may contain hundreds to thousands of triangles. Such a large data vector, and consequently covariance matrix, makes it very tricky to use the numerical covariance matrix as defined in Eq.\,\eqref{eq:defnumcov}. This is because we usually need $N_{\rm sims} \gg N $ to beat numerical noise, $N_{\rm sims}$ being the number of simulations and $N$ the dimension of the data vector.  Moreover, whenever the number of simulations is lower than the dimension of the data vector, the numerical covariance matrix cannot be inverted. Secondly, it is a way of testing a more realistic scenario, as the modeling of the bispectrum is significantly more involved than the one for a two-point function. In fact, as we will see, the model covariance we are going to use is knowingly incomplete, and our method will not be able to perform as well as a numerical covariance matrix drawn from $N_{\rm sims} =10000$ simulations. It will show, however, that a fit with parameters can help in building a reliable covariance matrix, even if the model is incomplete, as compared to considering a diagonal Gaussian covariance matrix.

\subsection{The bispectrum and its covariance}

Let us first set the notation on the bispectrum. Differently than the previous example, we work in Fourier space, and the bispectrum is defined as
\begin{equation}
    \langle \delta(\bk_1)\delta(\bk_2)\delta(\bk_3)\rangle = \frac{\delta_K(\bk_{1}+\bk_{2}+\bk_{3})}{k^3_f} B(k_1,k_2,k_3),
\end{equation}
where $\delta(\bk)$ is the discrete Fourier transform of the density contrast $\delta(\bx)$, $\delta_K$ is the Kronecker symbol (equal to unity when the argument vanishes, zero otherwise) and $k_f=2\pi/L$ is the fundamental frequency of a cubic box of volume $L^3$. 
We measure the bispectrum for all the $10000$ Pinocchio halo catalogs. We use unbiased estimators for the measurement of the bispectrum following the definition of \cite{ScoccimarroEtal1998, Scoccimarro2015}. We implement a fourth-order density interpolation and the interlacing scheme described in \cite{SefusattiEtal2016}. We divide $k$-modes into bins of width $\Delta k = k_f$ and up to a $k_{\rm max}=0.12$ $h/$Mpc, for a total of $29$ $k$-bins and $2766$ triangles.

\begin{figure}
    \centering
    \includegraphics[width=0.47\textwidth]{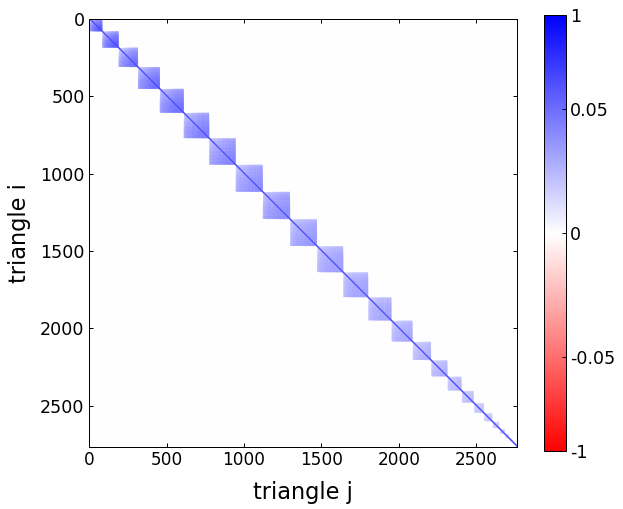}
    \includegraphics[width=0.47\textwidth]{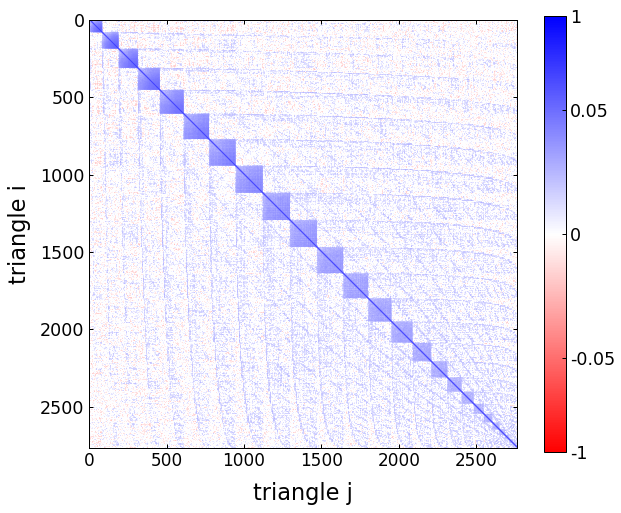}   
    \caption{Cross-correlation matrix $r_{ij}$ of the halo bispectrum covariance for the model covariance (left) and the numerical covariance computed on $N_{\rm sims} =10000$ simulations (right).}
    \label{fig_13}
\end{figure}

We consider the approximation for the bispectrum covariance given by\footnote{The complete formula for the bispectrum covariance would read
\begin{align}\label{eq:biscov}
C^{B}_{ij} &= C^{B, (PPP)}_{ij} + C^{B, (BB)}_{ij} + C^{B, (PT)}_{ij} + C^{B, (P_6)}_{ij}\,,
\end{align}
where $ T$ and $P_6$ are the trispectrum and pentaspectrum, respectively. The ``PT'' term represents the product of the power spectrum and trispectrum. As explained below, we are approximating the PT term to be proportional to the BB term and neglect $P_6$. See \cite{Biagetti:2021tua} for more details.} \cite{SefusattiEtal2006, BiagettiEtal2021A}
\begin{align}\label{eq:bmodel}
    C^{B}_{ij} &= \alpha\,C^{B, (PPP)}_{ij} + \beta\,\delta^K_{k_3^ik_3^j} \,C^{B, (BB)}_{ij},
\end{align}
where $\alpha$ and $\beta$ are free parameters  and $\delta^K_{k_3^ik_3^j}$ is a Kronecker symbol which is non-vanishing when the two triangles $i$ and $j$ have the smallest side in common, i.e. $k_3^i$ is equal to $k_3^j$. The first term, $C^{B, (PPP)}_{ij}$, is the Gaussian contribution, which we write down in the thin-shell approximation as
\begin{equation}
\label{eq:CBB-PPP}
C^{B,(PPP)}_{ij} 
 \simeq \frac{\delta_{ij}\,s_B}{k_f^3 N_{tr}^i}\, P(k_1^i)\,P(k_2^i)\,P(k_3^i)\,, 
\end{equation}
where $s_B = 6, 2, 1$ is the symmetry factor accounting for the shape of the triangles (equilateral, isosceles and scalene, respectively) and $N_{tr}^i$ is the number of fundamental triangles in the triangle bin $\left\{k_1^i, k_2^i, k_3^i\right\}$. The second term is defined as
\begin{equation}
    C^{B,(BB)}_{ij}  \simeq B_i\,B_j\,\left(\Sigma^{11}_{ij} + 8~{\rm perm.} \right)\label{appeqCBB-BB} \,,\\
\end{equation}
where $i$ and $j$ indicate the triangle bins $\{k_1^i, k_2^i, k_3^i\}$ and $\{k_1^j, k_2^j, k_3^j\}$, respectively, $B_i$ is the bispectrum for the triangle bin $i$ and  $\Sigma^{ab}_{ij}$ is a mode-counting factor that depends again on the shape of the triangle. Both terms are computed using measurements of the power spectrum and bispectrum directly, without using any perturbative calculation. The resulting covariance has a block-diagonal structure, which we show in the left panel of Fig.\,\ref{fig_13} by plotting the cross-correlation matrix
\begin{equation}
    r_{ij}=\frac{C_{ij}}{\sqrt{C_{ii}C_{jj}}},
\end{equation}
which helps in visualizing the importance of off-diagonal elements with respect to diagonal ones.

For comparison, we also plot the numerical covariance for $N_{\rm sims} =10000$, in the right panel of the same figure. The numerical covariance shows a similar block diagonal structure as the modeled one. Indeed, these particular non-diagonal entries of the bispectrum covariance are the largest terms in the non-Gaussian covariance \cite{BiagettiEtal2021A}. Based on theoretical considerations \cite{Barreira2019, BiagettiEtal2021A}, we expect that $\alpha \simeq 1$ and $\beta \simeq 2$. In particular, $\beta=2$ assumes that the contribution due the product of the power spectrum and trispectrum of the field can be approximated by the term in Eq.~\eqref{appeqCBB-BB}, a good approximation for squeezed triangles, but not for generic shapes. In the numerical covariance, we can clearly see more structure outside the blocks, which Eq.\,\eqref{eq:bmodel} does not model.

\subsection{Results}

Having defined the model covariance for the bispectrum test, we can proceed with the same steps as done for the two-point correlation function in the  Section \ref{sec:test1}. 
 \begin{figure}
    \centering
    \includegraphics[width=0.55\textwidth]{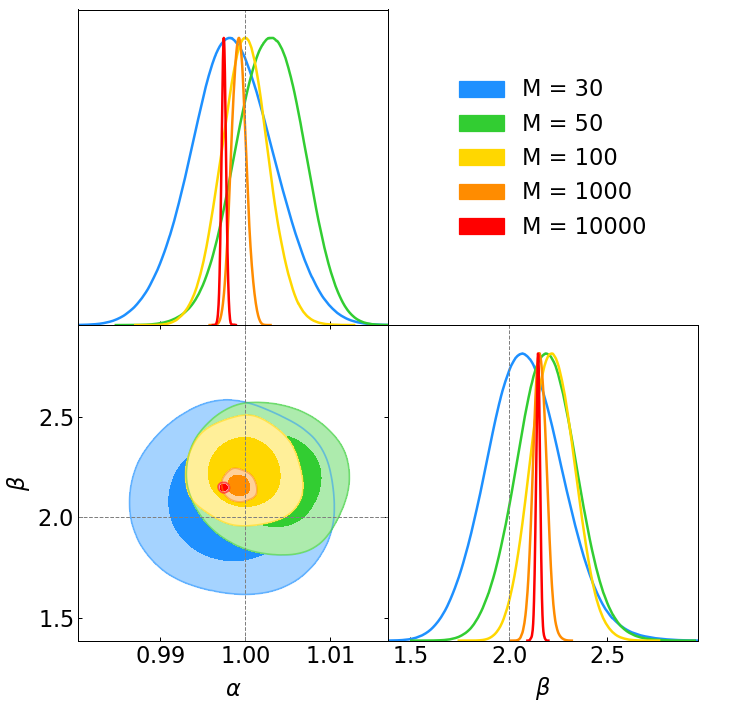}   
    \caption{Same of left panel of Fig.\,\ref{fig_fit} for the bispectrum covariance. The parameters $\alpha$ and $\beta$ represent the amplitude of the Gaussian and (part of the) non-Gaussian covariance, respectively, as defined in Eq.\,\eqref{eq:bmodel}. }
    \label{fig_10}
\end{figure}
\subsubsection{Maximizing the likelihood}

We maximize the likelihood of Eq.~(\ref{eq:master}) varying $\alpha$ and $\beta$ using sets of $C_n$ for $N_{\rm sims} =30$, $50$, $100$, $1000$ and $10000$ as in the previous test. We plot the contour plots for $\alpha$ and $\beta$ in Fig.\,\ref{fig_10}. Interestingly, the theoretical values do not fall within the contours for $N_{\rm sims} =1000$ and $N_{\rm sims} =10000$. In fact, this is not surprising, since the model covariance is an incomplete model of the full covariance, which has non-zero elements also outside the block-diagonal structure modeled by Eq.\,\eqref{eq:bmodel}. These non-zero elements might be sourced by a connected 6-point function, or by correlated noise. Even though these terms are small compared to the ones we model, when considering a large number of simulations they become significant with respect to sample noise and the fit tries to adapt to them shifting the central values of $\alpha$ and $\beta$.

\begin{figure}
    \centering
    \includegraphics[width=1.05\textwidth]{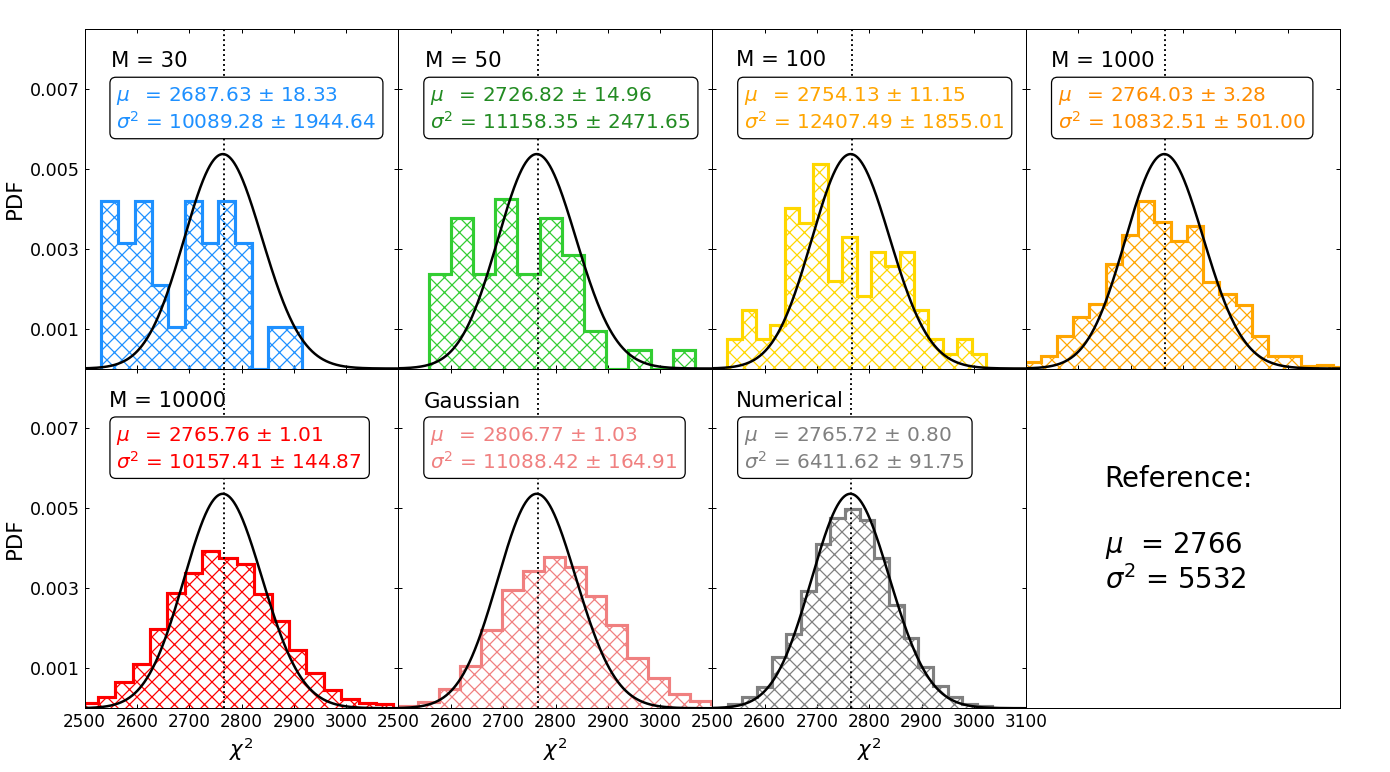}   
    \caption{Same of Fig.\,\ref{fig_chi2}, for the bispectrum covariance.}
    \label{fig_11}
\end{figure}

\subsubsection{A $\chi^2$ test for the inverse of the covariance}

Using the best-fit values for the $\alpha$ and $\beta$ parameters obtained from maximizing the likelihood, we perform the $\chi^2$ test introduced in the previous section in order to verify the performance of our model covariance. We show results in Fig.\,\ref{fig_11}. Differently than for the two-point correlation function, our model for the bispectrum covariance does not perform as well as the numerical covariance obtained from $N_{\rm sims} =10000$ simulations. Indeed, even though the mean $\mu$ of the $\chi^2$ distribution of our model covariance is in good agreement with the theoretical value of $\mu=2766$ already at $N_{\rm sims} =100$ simulations, the variance $\sigma^2$ is significantly off.\footnote{It is interesting to point out that even the numerical covariance does not fit perfectly well the reference $\chi^2$. We attribute this fact to the actual likelihood of the bispectrum not being Gaussian, at this volume.} Again, we can expect this given that our model covariance is incomplete. Nevertheless, it is useful to point out that, in the absence of a theoretical prior on the free parameters, and a numerical covariance, this method would prove useful in building a good approximation of the true covariance. Indeed, given that the mean of the $\chi^2$ is within the theoretical value, we expect that our model covariance does not bias strongly parameter estimation. For comparison, we compute the histogram also for the pure Gaussian covariance, i.e. a model covariance built fixing $\alpha=1$ and $\beta=0$, see Eq.~\eqref{eq:CBB-PPP}. In this case, both the mean and the variance of the histogram are significantly off from the theoretical curve.

\begin{figure}
    \centering
    \includegraphics[width=0.55\textwidth]{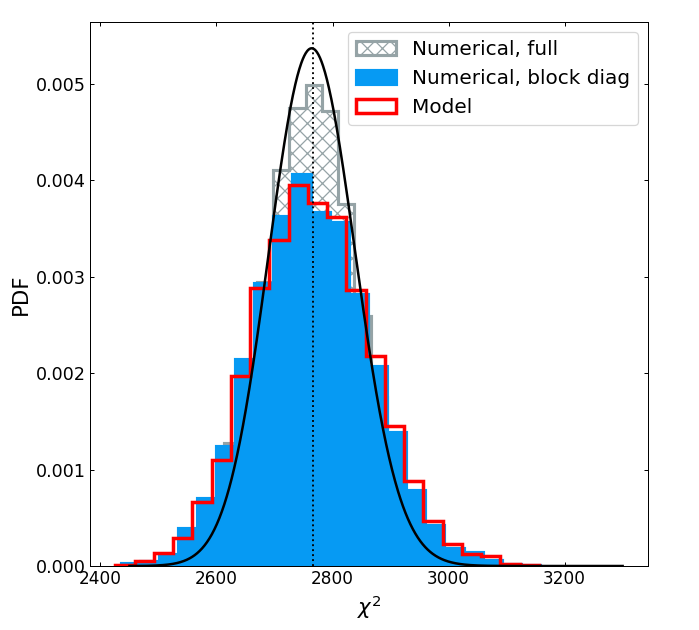}   
    \caption{$\chi^2$ distribution for the full numerical (gray), block-diagonal numerical (blue), and model (red) covariance.}
    \label{fig_12}
\end{figure}

In order to further confirm that the failure of the $\chi^2$ test is linked to the incompleteness of the model, we perform the following test: we calculate the numerical covariance matrix where we only keep the block diagonals that are modeled by our model covariance, putting all other off-diagonal terms to zero. We then redo the $\chi^2$ test using this numerical covariance and compare to the result that we got using the model covariance with $N_{\rm sims} =10000$ simulations. We show the result of this test in Fig.\,\ref{fig_12}. Indeed, if we only consider the blocks that we model in the numerical covariance, the $\chi^2$ test gives a very similar result as for the model covariance.

\section{Conclusions}

In this paper, we have explored a two-steps method to build a reliable and cheap covariance matrix. It is based on using two basic ingredients: an even imperfect model covariance with free parameters, and a $\chi^2$-test. Upon successfully finding the best-fit values for the free-parameters for which the $\chi^2$-test is passed, the methods provides a reliable covariance matrix having to run a small number of simulations, typically smaller than the data-vector itself. We have applied the method to two contexts, using the two-point correlation function and the bispectrum of mock halo catalogs as observables. We employed knowingly incomplete models of the covariance with two free parameters in both cases. We have shown how to verify their reliability without relying on knowing the true covariance via a $\chi^2$ test. In the case of the two-point correlation function, using only $100$ simulations, we are able to recover unbiased estimates of the cosmological parameters of the simulation.  We found that our covariance matrix fit prefers a non-Poisson shot noise in our simple  model. Since the effective shot noise is different when derived from covariance matrix compared to direct power spectrum fit, the most likely explanation is that we are detecting higher-order corrections. In the second example, using a non-Gaussian model for the bispectrum covariance we improve significantly with respect to a Gaussian covariance, but we are not able to match the numerical covariance drawn from $N_{\rm sims} =10000$.

Consistent with our analytic results in Section \ref{sec:chi2test} we find in both cases that the first moment of $\chi^2$ is consistent with theoretical expectation. The second moment is correct for the case of correlation function, but too high in the case of bispectrum. There the wrong covariance matrix causes an extra excess on the $\chi^2$ values with rms of $\sim 60$, similar to intrinsic scatter from $\chi^2$ distribution.

The main strength of the method is that it can be applied to test a proposed covariance matrix even in the case where the number of simulated universes is not sufficient to generate even a positive-definite matrix. It relies on examining the consistency of $\chi^2$ values derived from a model covariance matrix with theoretical expectation. A simple way of doing this is to measure mean and variance of realization $\chi^2$ values, comparing these with expectation values for a $\chi^2$ distribution. Even when the test is performed on the same simulated realization used for fitting the covariance matrix, the result might be inconsistent if the covariance model is insufficiently flexible to describe the actual covariance. The results also quantify the badness of the covariance matrix by giving typical expected biases in $\chi^2$ values.

We also found that the adopted model can have intrinsic biases when fitting an insufficient number of simulations. For example, when fitting with only 30 simulations, the derived best-fit covariance matrix not only varies with sample variance but is systematically underestimated (see Fig.\,\ref{fig_fit}) for this subset of simulations. 
Investigating this effect further and developing covariance models that are unbiased (in the sense that they in average produce an unbiased covariance matrix) is left to future work. 

In practice we have found that even covariance matrices that have a demonstrably biased $\chi^2$ distributions often perform well enough in practical situations, giving cosmological parameters constraints that are acceptably biased with respect to the ideal case (See Fig.\,\ref{fig_6}). This indicates that while our method allows for a proper propagation of covariance matrix uncertainties coming from sample variance on the number of realizations used into cosmological inferences, this might very rarely be used in practice. At the same time, the distribution of simulated $\chi^2$ values might allow us to ``recalibrate'' the goodness of fit measures. In our bispectrum example, the fitted covariance matrix produces a distribution of $\chi^2$ values which is too broad. This would allow us to better quantify the goodness of fit on the real data, which might be formally bad, but consistent with the distribution obtained with simulation with an imperfect covariance matrix. 

While our test shows in principle the potential of our method, there are several more realistic setups where this method could prove to be crucial. For instance, it will be important to further test to what degree the fitting procedure can compensate for an incomplete modeling of the covariance and if approximate phenomenological terms can be added into covariance matrix model that absorb terms missing from the theory. 
A good testing ground for these tests is the galaxy/halo bispectrum framework we introduced in Section \ref{sec:test2}, since the covariance in this case has large off-diagonal terms for which we have only an incomplete model.
We leave these tests to future work.

\section*{Acknowledgments}

M.B acknowledges support from the Netherlands Organization for Scientific Research (NWO), which is funded by the Dutch Ministry of Education, Culture and Science (OCW) under VENI grant 016.Veni.192.210. E.S. and P.M. are partially supported by the INFN INDARK PD51 grant. A.S. acknowledges hospitality of Institute for Fundamental Physics of the Universe in Trieste where this work was initiated. A.F. and A.S. are supported by the ERC-StG ‘ClustersXCosmo’ grant agreement 716762, by the FARE-MIUR grant 'ClustersXEuclid' R165SBKTMA, and by INFN InDark Grant.

\bibliographystyle{JHEP}
\bibliography{cosmologia,references}

\providecommand{\href}[2]{#2}\begingroup\raggedright\begin{thebibliography}{10}

\bibitem{BlakeEtal2011}
C.~{Blake}, S.~{Brough}, M.~{Colless}, C.~{Contreras}, W.~{Couch}, S.~{Croom}
  et~al., \emph{{The WiggleZ Dark Energy Survey: the growth rate of cosmic
  structure since redshift z=0.9}},
  \href{https://doi.org/10.1111/j.1365-2966.2011.18903.x}{\emph{\mnras}
  {\bfseries 415} (2011) 2876}
  [\href{https://arxiv.org/abs/1104.2948}{{\ttfamily 1104.2948}}].

\bibitem{DelaTorreEtal2013}
S.~{de la Torre}, L.~{Guzzo}, J.A.~{Peacock}, E.~{Branchini}, A.~{Iovino},
  B.R.~{Granett} et~al., \emph{{The VIMOS Public Extragalactic Redshift Survey
  (VIPERS) . Galaxy clustering and redshift-space distortions at z {$\sim$} 0.8
  in the first data release}},
  \href{https://doi.org/10.1051/0004-6361/201321463}{\emph{\aap} {\bfseries
  557} (2013) A54} [\href{https://arxiv.org/abs/1303.2622}{{\ttfamily
  1303.2622}}].

\bibitem{AlamEtal2017}
S.~{Alam}, M.~{Ata}, S.~{Bailey}, F.~{Beutler}, D.~{Bizyaev}, J.A.~{Blazek}
  et~al., \emph{{The clustering of galaxies in the completed SDSS-III Baryon
  Oscillation Spectroscopic Survey: cosmological analysis of the DR12 galaxy
  sample}}, \href{https://doi.org/10.1093/mnras/stx721}{\emph{\mnras}
  {\bfseries 470} (2017) 2617}
  [\href{https://arxiv.org/abs/1607.03155}{{\ttfamily 1607.03155}}].

\bibitem{GilMarinEtal2017}
H.~{Gil-Mar{\'{\i}}n}, W.J.~{Percival}, L.~{Verde}, J.R.~{Brownstein},
  C.-H.~{Chuang}, F.-S.~{Kitaura} et~al., \emph{{The clustering of galaxies in
  the SDSS-III Baryon Oscillation Spectroscopic Survey: RSD measurement from
  the power spectrum and bispectrum of the DR12 BOSS galaxies}},
  \href{https://doi.org/10.1093/mnras/stw2679}{\emph{\mnras} {\bfseries 465}
  (2017) 1757} [\href{https://arxiv.org/abs/1606.00439}{{\ttfamily
  1606.00439}}].

\bibitem{AsgariEtal2021}
M.~{Asgari}, C.-A.~{Lin}, B.~{Joachimi}, B.~{Giblin}, C.~{Heymans},
  H.~{Hildebrandt} et~al., \emph{{KiDS-1000 cosmology: Cosmic shear constraints
  and comparison between two point statistics}},
  \href{https://doi.org/10.1051/0004-6361/202039070}{\emph{\aap} {\bfseries
  645} (2021) A104} [\href{https://arxiv.org/abs/2007.15633}{{\ttfamily
  2007.15633}}].

\bibitem{AbbottEtal2022}
T.M.C.~{Abbott}, M.~{Aguena}, A.~{Alarcon}, S.~{Allam}, O.~{Alves}, A.~{Amon}
  et~al., \emph{{Dark Energy Survey Year 3 results: Cosmological constraints
  from galaxy clustering and weak lensing}},
  \href{https://doi.org/10.1103/PhysRevD.105.023520}{\emph{\prd} {\bfseries
  105} (2022) 023520} [\href{https://arxiv.org/abs/2105.13549}{{\ttfamily
  2105.13549}}].

\bibitem{PhilcoxIvanov2022}
O.H.E.~{Philcox} and M.M.~{Ivanov}, \emph{{BOSS DR12 full-shape cosmology:
  {\ensuremath{\Lambda}} CDM constraints from the large-scale galaxy power
  spectrum and bispectrum monopole}},
  \href{https://doi.org/10.1103/PhysRevD.105.043517}{\emph{\prd} {\bfseries
  105} (2022) 043517} [\href{https://arxiv.org/abs/2112.04515}{{\ttfamily
  2112.04515}}].

\bibitem{DAmicoEtal2020}
G.~{d'Amico}, J.~{Gleyzes}, N.~{Kokron}, K.~{Markovic}, L.~{Senatore},
  P.~{Zhang} et~al., \emph{{The cosmological analysis of the SDSS/BOSS data
  from the Effective Field Theory of Large-Scale Structure}},
  \href{https://doi.org/10.1088/1475-7516/2020/05/005}{\emph{\jcap} {\bfseries
  2020} (2020) 005} [\href{https://arxiv.org/abs/1909.05271}{{\ttfamily
  1909.05271}}].

\bibitem{Biagetti:2020skr}
M.~Biagetti, A.~Cole and G.~Shiu, \emph{{The Persistence of Large Scale
  Structures I: Primordial non-Gaussianity}},
  \href{https://doi.org/10.1088/1475-7516/2021/04/061}{\emph{JCAP} {\bfseries
  04} (2021) 061} [\href{https://arxiv.org/abs/2009.04819}{{\ttfamily
  2009.04819}}].

\bibitem{Heydenreich:2020hrr}
S.~Heydenreich, B.~Br\"uck and J.~Harnois-D\'eraps, \emph{{Persistent homology
  in cosmic shear: constraining parameters with topological data analysis}},
  \href{https://doi.org/10.1051/0004-6361/202039048}{\emph{Astron. Astrophys.}
  {\bfseries 648} (2021) A74}
  [\href{https://arxiv.org/abs/2007.13724}{{\ttfamily 2007.13724}}].

\bibitem{BiagettiEtal2022A}
M.~{Biagetti}, J.~{Calles}, L.~{Castiblanco}, A.~{Cole} and J.~{Nore{\~n}a},
  \emph{{Fisher Forecasts for Primordial non-Gaussianity from Persistent
  Homology}}, {\emph{arXiv e-prints} (2022) arXiv:2203.08262}
  [\href{https://arxiv.org/abs/2203.08262}{{\ttfamily 2203.08262}}].

\bibitem{Heydenreich:2022dci}
S.~Heydenreich, B.~Br\"uck, P.~Burger, J.~Harnois-D\'eraps, S.~Unruh, T.~Castro
  et~al., \emph{{Persistent homology in cosmic shear II: A tomographic analysis
  of DES-Y1}},  \href{https://arxiv.org/abs/2204.11831}{{\ttfamily
  2204.11831}}.

\bibitem{2010MNRAS.403.1392L}
G.~{Lavaux} and B.D.~{Wandelt}, \emph{{Precision cosmology with voids:
  definition, methods, dynamics}},
  \href{https://doi.org/10.1111/j.1365-2966.2010.16197.x}{\emph{\mnras}
  {\bfseries 403} (2010) 1392}
  [\href{https://arxiv.org/abs/0906.4101}{{\ttfamily 0906.4101}}].

\bibitem{2006MNRAS.369...68S}
R.~{Skibba}, R.K.~{Sheth}, A.J.~{Connolly} and R.~{Scranton}, \emph{{The
  luminosity-weighted or `marked' correlation function}},
  \href{https://doi.org/10.1111/j.1365-2966.2006.10196.x}{\emph{\mnras}
  {\bfseries 369} (2006) 68}
  [\href{https://arxiv.org/abs/astro-ph/0512463}{{\ttfamily
  astro-ph/0512463}}].

\bibitem{2016JCAP...11..057W}
M.~{White}, \emph{{A marked correlation function for constraining modified
  gravity models}},
  \href{https://doi.org/10.1088/1475-7516/2016/11/057}{\emph{\jcap} {\bfseries
  2016} (2016) 057} [\href{https://arxiv.org/abs/1609.08632}{{\ttfamily
  1609.08632}}].

\bibitem{2018MNRAS.478.3627A}
J.~{Armijo}, Y.-C.~{Cai}, N.~{Padilla}, B.~{Li} and J.A.~{Peacock},
  \emph{{Testing modified gravity using a marked correlation function}},
  \href{https://doi.org/10.1093/mnras/sty1335}{\emph{\mnras} {\bfseries 478}
  (2018) 3627} [\href{https://arxiv.org/abs/1801.08975}{{\ttfamily
  1801.08975}}].

\bibitem{ManeraEtal2013}
M.~Manera, R.~Scoccimarro, W.J.~Percival, L.~{Samushia}, C.K.~McBride,
  A.J.~Ross et~al., \emph{{The clustering of galaxies in the SDSS-III Baryon
  Oscillation Spectroscopic Survey: a large sample of mock galaxy catalogues}},
  \href{https://doi.org/10.1093/mnras/sts084}{\emph{\mnras} {\bfseries 428}
  (2013) 1036} [\href{https://arxiv.org/abs/1203.6609}{{\ttfamily 1203.6609}}].

\bibitem{KitauraEtal2016}
F.-S.~{Kitaura}, S.~{Rodr{\'{\i}}guez-Torres}, C.-H.~{Chuang}, C.~{Zhao},
  F.~{Prada}, H.~{Gil-Mar{\'{\i}}n} et~al., \emph{{The clustering of galaxies
  in the SDSS-III Baryon Oscillation Spectroscopic Survey: mock galaxy
  catalogues for the BOSS Final Data Release}},
  \href{https://doi.org/10.1093/mnras/stv2826}{\emph{\mnras} {\bfseries 456}
  (2016) 4156} [\href{https://arxiv.org/abs/1509.06400}{{\ttfamily
  1509.06400}}].

\bibitem{AvilaEtal2018}
S.~{Avila}, M.~{Crocce}, A.J.~{Ross}, J.~{Garc{\'\i}a-Bellido},
  W.J.~{Percival}, N.~{Banik} et~al., \emph{{Dark Energy Survey Year-1 results:
  galaxy mock catalogues for BAO}},
  \href{https://doi.org/10.1093/mnras/sty1389}{\emph{\mnras} {\bfseries 479}
  (2018) 94} [\href{https://arxiv.org/abs/1712.06232}{{\ttfamily 1712.06232}}].

\bibitem{Monaco2016}
P.~{Monaco}, \emph{{Approximate Methods for the Generation of Dark Matter Halo
  Catalogs in the Age of Precision Cosmology}},
  \href{https://doi.org/10.3390/galaxies4040053}{\emph{Galaxies} {\bfseries 4}
  (2016) 53} [\href{https://arxiv.org/abs/1605.07752}{{\ttfamily 1605.07752}}].

\bibitem{FengEtal2016}
Y.~Feng, M.-Y.~Chu, U.~Seljak and P.~McDonald, \emph{{FASTPM: a new scheme for
  fast simulations of dark matter and haloes}},
  \href{https://doi.org/10.1093/mnras/stw2123}{\emph{\mnras} {\bfseries 463}
  (2016) 2273} [\href{https://arxiv.org/abs/1603.00476}{{\ttfamily
  1603.00476}}].

\bibitem{LippichEtal2019}
M.~{Lippich}, A.G.~{S{\'a}nchez}, M.~{Colavincenzo}, E.~{Sefusatti},
  P.~{Monaco}, L.~{Blot} et~al., \emph{{Comparing approximate methods for mock
  catalogues and covariance matrices - I. Correlation function}},
  \href{https://doi.org/10.1093/mnras/sty2757}{\emph{\mnras} {\bfseries 482}
  (2019) 1786} [\href{https://arxiv.org/abs/1806.09477}{{\ttfamily
  1806.09477}}].

\bibitem{BlotEtal2019}
L.~{Blot}, M.~{Crocce}, E.~{Sefusatti}, M.~{Lippich}, A.G.~{S{\'a}nchez},
  M.~{Colavincenzo} et~al., \emph{{Comparing approximate methods for mock
  catalogues and covariance matrices II: Power spectrum multipoles}},
  \href{https://doi.org/10.1093/mnras/stz507}{\emph{\mnras} (2019) }
  [\href{https://arxiv.org/abs/1806.09497}{{\ttfamily 1806.09497}}].

\bibitem{ColavincenzoEtal2019}
M.~{Colavincenzo}, E.~{Sefusatti}, P.~{Monaco}, L.~{Blot}, M.~{Crocce},
  M.~{Lippich} et~al., \emph{{Comparing approximate methods for mock catalogues
  and covariance matrices - III: bispectrum}},
  \href{https://doi.org/10.1093/mnras/sty2964}{\emph{\mnras} {\bfseries 482}
  (2019) 4883} [\href{https://arxiv.org/abs/1806.09499}{{\ttfamily
  1806.09499}}].

\bibitem{HartlapEtal2009}
J.~{Hartlap}, T.~{Schrabback}, P.~{Simon} and P.~{Schneider}, \emph{{The
  non-Gaussianity of the cosmic shear likelihood or how odd is the Chandra Deep
  Field South?}},
  \href{https://doi.org/10.1051/0004-6361/200911697}{\emph{\aap} {\bfseries
  504} (2009) 689} [\href{https://arxiv.org/abs/0901.3269}{{\ttfamily
  0901.3269}}].

\bibitem{TaylorJoachimiKitching2013}
A.N.~Taylor, B.~Joachimi and T.D.~{Kitching}, \emph{{Putting the precision in
  precision cosmology: How accurate should your data covariance matrix be?}},
  \href{https://doi.org/10.1093/mnras/stt270}{\emph{\mnras} {\bfseries 432}
  (2013) 1928} [\href{https://arxiv.org/abs/1212.4359}{{\ttfamily 1212.4359}}].

\bibitem{DodelsonSchneider2013}
S.~{Dodelson} and M.D.~{Schneider}, \emph{{The effect of covariance estimator
  error on cosmological parameter constraints}},
  \href{https://doi.org/10.1103/PhysRevD.88.063537}{\emph{\prd} {\bfseries 88}
  (2013) 063537} [\href{https://arxiv.org/abs/1304.2593}{{\ttfamily
  1304.2593}}].

\bibitem{PercivalEtal2014}
W.J.~{Percival}, A.J.~{Ross}, A.G.~{S{\'a}nchez}, L.~{Samushia}, A.~{Burden},
  R.~{Crittenden} et~al., \emph{{The clustering of Galaxies in the SDSS-III
  Baryon Oscillation Spectroscopic Survey: including covariance matrix
  errors}}, \href{https://doi.org/10.1093/mnras/stu112}{\emph{\mnras}
  {\bfseries 439} (2014) 2531}
  [\href{https://arxiv.org/abs/1312.4841}{{\ttfamily 1312.4841}}].

\bibitem{SellentinHeavens2016}
E.~{Sellentin} and A.F.~{Heavens}, \emph{{Parameter inference with estimated
  covariance matrices}},
  \href{https://doi.org/10.1093/mnrasl/slv190}{\emph{\mnras} {\bfseries 456}
  (2016) L132} [\href{https://arxiv.org/abs/1511.05969}{{\ttfamily
  1511.05969}}].

\bibitem{Scoccimarro2000B}
R.~Scoccimarro, \emph{The bispectrum: From theory to observations},
  \href{https://doi.org/10.1086/317248}{\emph{\apj} {\bfseries 544} (2000) 597}
  [\href{https://arxiv.org/abs/astro-ph/0004086}{{\ttfamily
  astro-ph/0004086}}].

\bibitem{HamiltonRimesScoccimarro2006}
A.J.S.~Hamilton, C.D.~{Rimes} and R.~Scoccimarro, \emph{On measuring the
  covariance matrix of the non-linear power spectrum from simulations},
  \href{https://doi.org/10.1111/j.1365-2966.2006.10709.x}{\emph{\mnras}
  {\bfseries 371} (2006) 1188}
  [\href{https://arxiv.org/abs/arXiv:astro-ph/0511416}{{\ttfamily
  arXiv:astro-ph/0511416}}].

\bibitem{PopeSzapudi2008}
A.C.~Pope and I.~Szapudi, \emph{{Shrinkage estimation of the power spectrum
  covariance matrix}},
  \href{https://doi.org/10.1111/j.1365-2966.2008.13561.x}{\emph{\mnras}
  {\bfseries 389} (2008) 766}
  [\href{https://arxiv.org/abs/0711.2509}{{\ttfamily 0711.2509}}].

\bibitem{Joachimi2017}
B.~{Joachimi}, \emph{{Non-linear shrinkage estimation of large-scale structure
  covariance}}, \href{https://doi.org/10.1093/mnrasl/slw240}{\emph{\mnras}
  {\bfseries 466} (2017) L83}
  [\href{https://arxiv.org/abs/1612.00752}{{\ttfamily 1612.00752}}].

\bibitem{PazSanchez2015}
D.J.~{Paz} and A.G.~S{\'a}nchez, \emph{{Improving the precision matrix for
  precision cosmology}},
  \href{https://doi.org/10.1093/mnras/stv2259}{\emph{\mnras} {\bfseries 454}
  (2015) 4326} [\href{https://arxiv.org/abs/1508.03162}{{\ttfamily
  1508.03162}}].

\bibitem{ChartierWandelt2022}
N.~{Chartier} and B.D.~{Wandelt}, \emph{{CARPool covariance: fast, unbiased
  covariance estimation for large-scale structure observables}},
  \href{https://doi.org/10.1093/mnras/stab3097}{\emph{\mnras} {\bfseries 509}
  (2022) 2220} [\href{https://arxiv.org/abs/2106.11718}{{\ttfamily
  2106.11718}}].

\bibitem{deSanti:2022nlz}
N.S.M.~de~Santi and L.R.~Abramo, \emph{{Improving cosmological covariance
  matrices with machine learning}},
  \href{https://arxiv.org/abs/2205.10881}{{\ttfamily 2205.10881}}.

\bibitem{Lacasa2020}
F.~{Lacasa}, \emph{{The impact of braiding covariance and in-survey covariance
  on next-generation galaxy surveys}},
  \href{https://doi.org/10.1051/0004-6361/201936683}{\emph{\aap} {\bfseries
  634} (2020) A74} [\href{https://arxiv.org/abs/1909.00791}{{\ttfamily
  1909.00791}}].

\bibitem{FangEiflerKrause2020}
X.~{Fang}, T.~{Eifler} and E.~{Krause}, \emph{{2D-FFTLog: efficient computation
  of real-space covariance matrices for galaxy clustering and weak lensing}},
  \href{https://doi.org/10.1093/mnras/staa1726}{\emph{\mnras} {\bfseries 497}
  (2020) 2699} [\href{https://arxiv.org/abs/2004.04833}{{\ttfamily
  2004.04833}}].

\bibitem{SugiyamaEtal2020}
N.S.~{Sugiyama}, S.~{Saito}, F.~{Beutler} and H.-J.~{Seo}, \emph{{Perturbation
  theory approach to predict the covariance matrices of the galaxy power
  spectrum and bispectrum in redshift space}},
  \href{https://doi.org/10.1093/mnras/staa1940}{\emph{\mnras} {\bfseries 497}
  (2020) 1684} [\href{https://arxiv.org/abs/1908.06234}{{\ttfamily
  1908.06234}}].

\bibitem{WadekarScoccimarro2020}
D.~{Wadekar} and R.~{Scoccimarro}, \emph{{Galaxy power spectrum multipoles
  covariance in perturbation theory}},
  \href{https://doi.org/10.1103/PhysRevD.102.123517}{\emph{\prd} {\bfseries
  102} (2020) 123517} [\href{https://arxiv.org/abs/1910.02914}{{\ttfamily
  1910.02914}}].

\bibitem{WadekarIvanovScoccimarro2020}
D.~{Wadekar}, M.M.~{Ivanov} and R.~{Scoccimarro}, \emph{{Cosmological
  constraints from BOSS with analytic covariance matrices}},
  \href{https://doi.org/10.1103/PhysRevD.102.123521}{\emph{\prd} {\bfseries
  102} (2020) 123521} [\href{https://arxiv.org/abs/2009.00622}{{\ttfamily
  2009.00622}}].

\bibitem{BiagettiEtal2021A}
M.~{Biagetti}, L.~{Castiblanco}, J.~{Nore{\~n}a} and E.~{Sefusatti}, \emph{{The
  Covariance of Squeezed Bispectrum Configurations}}, {\emph{arXiv e-prints}
  (2021) arXiv:2111.05887} [\href{https://arxiv.org/abs/2111.05887}{{\ttfamily
  2111.05887}}].

\bibitem{Xu:2012hg}
X.~Xu, N.~Padmanabhan, D.J.~Eisenstein, K.T.~Mehta and A.J.~Cuesta, \emph{{A
  2\% Distance to z=0.35 by Reconstructing Baryon Acoustic Oscillations - II:
  Fitting Techniques}},
  \href{https://doi.org/10.1111/j.1365-2966.2012.21573.x}{\emph{Mon. Not. Roy.
  Astron. Soc.} {\bfseries 427} (2012) 2146}
  [\href{https://arxiv.org/abs/1202.0091}{{\ttfamily 1202.0091}}].

\bibitem{OConnellEtal2016}
R.~{O'Connell}, D.J.~Eisenstein, M.~{Vargas}, S.~Ho and N.~Padmanabhan,
  \emph{{Large covariance matrices: smooth models from the two-point
  correlation function}},
  \href{https://doi.org/10.1093/mnras/stw1821}{\emph{\mnras} {\bfseries 462}
  (2016) 2681} [\href{https://arxiv.org/abs/1510.01740}{{\ttfamily
  1510.01740}}].

\bibitem{Slepian:2015hca}
Z.~Slepian et~al., \emph{{The large-scale 3-point correlation function of the
  SDSS BOSS DR12 CMASS galaxies}},
  \href{https://arxiv.org/abs/1512.02231}{{\ttfamily 1512.02231}}.

\bibitem{Pearson:2015gca}
D.W.~Pearson and L.~Samushia, \emph{{Estimating the power spectrum covariance
  matrix with fewer mock samples}},
  \href{https://doi.org/10.1093/mnras/stw062}{\emph{Mon. Not. Roy. Astron.
  Soc.} {\bfseries 457} (2016) 993}
  [\href{https://arxiv.org/abs/1509.00064}{{\ttfamily 1509.00064}}].

\bibitem{Hall:2018umb}
A.~Hall and A.~Taylor, \emph{{A Bayesian method for combining theoretical and
  simulated covariance matrices for large-scale structure surveys}},
  \href{https://doi.org/10.1093/mnras/sty3102}{\emph{Mon. Not. Roy. Astron.
  Soc.} {\bfseries 483} (2019) 189}
  [\href{https://arxiv.org/abs/1807.06875}{{\ttfamily 1807.06875}}].

\bibitem{DES:2020ypx}
{\scshape DES} collaboration, \emph{{Dark Energy Survey year 3 results:
  covariance modelling and its impact on parameter estimation and quality of
  fit}}, \href{https://doi.org/10.1093/mnras/stab2384}{\emph{Mon. Not. Roy.
  Astron. Soc.} {\bfseries 508} (2021) 3125}
  [\href{https://arxiv.org/abs/2012.08568}{{\ttfamily 2012.08568}}].

\bibitem{ScoccimarroZaldarriagaHui1999}
R.~Scoccimarro, M.~Zaldarriaga and L.~Hui, \emph{Power spectrum correlations
  induced by nonlinear clustering},
  \href{https://doi.org/10.1086/308059}{\emph{\apj} {\bfseries 527} (1999) 1}
  [\href{https://arxiv.org/abs/arXiv:astro-ph/9901099}{{\ttfamily
  arXiv:astro-ph/9901099}}].

\bibitem{GriebEtal2016}
J.N.~Grieb, A.G.~S{\'a}nchez, S.~Salazar-Albornoz and C.~Dalla~Vecchia,
  \emph{{Gaussian covariance matrices for anisotropic galaxy clustering
  measurements}}, \href{https://doi.org/10.1093/mnras/stw065}{\emph{\mnras}
  {\bfseries 457} (2016) 1577}
  [\href{https://arxiv.org/abs/1509.04293}{{\ttfamily 1509.04293}}].

\bibitem{MonacoTheunsTaffoni2002}
P.~Monaco, T.~{Theuns} and G.~{Taffoni}, \emph{{The pinocchio algorithm:
  pinpointing orbit-crossing collapsed hierarchical objects in a linear density
  field}},
  \href{https://doi.org/10.1046/j.1365-8711.2002.05162.x}{\emph{\mnras}
  {\bfseries 331} (2002) 587}
  [\href{https://arxiv.org/abs/arXiv:astro-ph/0109323}{{\ttfamily
  arXiv:astro-ph/0109323}}].

\bibitem{MonacoEtal2013}
P.~Monaco, E.~Sefusatti, S.~Borgani, M.~Crocce, P.~Fosalba, R.K.~Sheth et~al.,
  \emph{{An accurate tool for the fast generation of dark matter halo
  catalogues}}, \href{https://doi.org/10.1093/mnras/stt907}{\emph{\mnras}
  {\bfseries 433} (2013) 2389}
  [\href{https://arxiv.org/abs/1305.1505}{{\ttfamily 1305.1505}}].

\bibitem{MunariEtal2017}
E.~{Munari}, P.~{Monaco}, E.~{Sefusatti}, E.~{Castorina}, F.G.~{Mohammad},
  S.~{Anselmi} et~al., \emph{{Improving fast generation of halo catalogues with
  higher order Lagrangian perturbation theory}},
  \href{https://doi.org/10.1093/mnras/stw3085}{\emph{\mnras} {\bfseries 465}
  (2017) 4658} [\href{https://arxiv.org/abs/1605.04788}{{\ttfamily
  1605.04788}}].

\bibitem{OddoEtal2020}
A.~{Oddo}, E.~{Sefusatti}, C.~{Porciani}, P.~{Monaco} and A.G.~{S{\'a}nchez},
  \emph{{Toward a robust inference method for the galaxy bispectrum: likelihood
  function and model selection}},
  \href{https://doi.org/10.1088/1475-7516/2020/03/056}{\emph{\jcap} {\bfseries
  2020} (2020) 056} [\href{https://arxiv.org/abs/1908.01774}{{\ttfamily
  1908.01774}}].

\bibitem{OddoEtal2021}
A.~{Oddo}, F.~{Rizzo}, E.~{Sefusatti}, C.~{Porciani} and P.~{Monaco},
  \emph{{Cosmological parameters from the likelihood analysis of the galaxy
  power spectrum and bispectrum in real space}},
  \href{https://doi.org/10.1088/1475-7516/2021/11/038}{\emph{\jcap} {\bfseries
  2021} (2021) 038} [\href{https://arxiv.org/abs/2108.03204}{{\ttfamily
  2108.03204}}].

\bibitem{LandySzalay1993}
S.D.~Landy and A.S.~Szalay, \emph{Bias and variance of angular correlation
  functions}, \href{https://doi.org/10.1086/172900}{\emph{\apj} {\bfseries 412}
  (1993) 64}.

\bibitem{MarulliVeropalumboMoresco2016}
F.~{Marulli}, A.~{Veropalumbo} and M.~{Moresco}, \emph{{CosmoBolognaLib: C++
  libraries for cosmological calculations}},
  \href{https://doi.org/10.1016/j.ascom.2016.01.005}{\emph{Astronomy and
  Computing} {\bfseries 14} (2016) 35}
  [\href{https://arxiv.org/abs/1511.00012}{{\ttfamily 1511.00012}}].

\bibitem{PhilcoxEisenstein2019}
O.H.E.~{Philcox} and D.J.~{Eisenstein}, \emph{{Estimating covariance matrices
  for two- and three-point correlation function moments in Arbitrary Survey
  Geometries}}, \href{https://doi.org/10.1093/mnras/stz2896}{\emph{\mnras}
  {\bfseries 490} (2019) 5931}
  [\href{https://arxiv.org/abs/1910.04764}{{\ttfamily 1910.04764}}].

\bibitem{LiEtal2019}
Y.~{Li}, S.~{Singh}, B.~{Yu}, Y.~{Feng} and U.~{Seljak}, \emph{{Disconnected
  covariance of 2-point functions in large-scale structure}},
  \href{https://doi.org/10.1088/1475-7516/2019/01/016}{\emph{\jcap} {\bfseries
  2019} (2019) 016} [\href{https://arxiv.org/abs/1811.05714}{{\ttfamily
  1811.05714}}].

\bibitem{BuchnerEtal2014}
J.~{Buchner}, A.~{Georgakakis}, K.~{Nandra}, L.~{Hsu}, C.~{Rangel},
  M.~{Brightman} et~al., \emph{{X-ray spectral modelling of the AGN obscuring
  region in the CDFS: Bayesian model selection and catalogue}},
  \href{https://doi.org/10.1051/0004-6361/201322971}{\emph{\aap} {\bfseries
  564} (2014) A125} [\href{https://arxiv.org/abs/1402.0004}{{\ttfamily
  1402.0004}}].

\bibitem{AlbrechtEtal2006a}
A.~{Albrecht}, G.~{Bernstein}, R.~{Cahn}, W.L.~{Freedman}, J.~{Hewitt}, W.~{Hu}
  et~al., \emph{{Report of the Dark Energy Task Force}}, {\emph{arXiv e-prints}
  (2006) astro} [\href{https://arxiv.org/abs/astro-ph/0609591}{{\ttfamily
  astro-ph/0609591}}].

\bibitem{ScoccimarroEtal1998}
R.~Scoccimarro, S.~Colombi, J.N.~Fry, J.A.~Frieman, E.~Hivon and A.~Melott,
  \emph{Nonlinear evolution of the bispectrum of cosmological perturbations},
  \href{https://doi.org/10.1086/305399}{\emph{\apj} {\bfseries 496} (1998) 586}
  [\href{https://arxiv.org/abs/astro-ph/9704075}{{\ttfamily
  astro-ph/9704075}}].

\bibitem{Scoccimarro2015}
R.~{Scoccimarro}, \emph{{Fast estimators for redshift-space clustering}},
  \href{https://doi.org/10.1103/PhysRevD.92.083532}{\emph{\prd} {\bfseries 92}
  (2015) 083532} [\href{https://arxiv.org/abs/1506.02729}{{\ttfamily
  1506.02729}}].

\bibitem{SefusattiEtal2016}
E.~{Sefusatti}, M.~{Crocce}, R.~{Scoccimarro} and H.M.P.~{Couchman},
  \emph{{Accurate estimators of correlation functions in Fourier space}},
  \href{https://doi.org/10.1093/mnras/stw1229}{\emph{\mnras} {\bfseries 460}
  (2016) 3624} [\href{https://arxiv.org/abs/1512.07295}{{\ttfamily
  1512.07295}}].

\bibitem{Biagetti:2021tua}
M.~Biagetti, L.~Castiblanco, J.~Nore\~na and E.~Sefusatti, \emph{{The
  Covariance of Squeezed Bispectrum Configurations}},
  \href{https://arxiv.org/abs/2111.05887}{{\ttfamily 2111.05887}}.

\bibitem{SefusattiEtal2006}
E.~Sefusatti, M.~Crocce, S.~Pueblas and R.~Scoccimarro, \emph{Cosmology and the
  bispectrum}, \href{https://doi.org/10.1103/PhysRevD.74.023522}{\emph{\prd}
  {\bfseries 74} (2006) 023522} [\href{https://arxiv.org/abs/arXiv:
  astro-ph/0604505}{{\ttfamily arXiv: astro-ph/0604505}}].

\bibitem{Barreira2019}
A.~{Barreira}, \emph{{The squeezed matter bispectrum covariance with
  responses}},
  \href{https://doi.org/10.1088/1475-7516/2019/03/008}{\emph{\jcap} {\bfseries
  2019} (2019) 008} [\href{https://arxiv.org/abs/1901.01243}{{\ttfamily
  1901.01243}}].

\end{thebibliography}\endgroup

\end{document}